\documentclass[journal,twoside]{ieeeaccess}
\usepackage{cite}
\usepackage{amsmath,amssymb,amsfonts}
\usepackage{algorithmic}
\usepackage{graphicx}
\usepackage{textcomp}

\usepackage{xr}
\usepackage{amsfonts,amsmath,amsthm,amssymb}
\usepackage{url,cite,bm,bbm, dsfont}
\usepackage{graphicx,xspace,epsfig,syntonly,dsfont}
\usepackage{xcolor}
\usepackage{siunitx}
\usepackage[stable]{footmisc}
\ifCLASSINFOpdf
\usepackage{optidef}
\usepackage{cleveref}
\usepackage{makecell}
\usepackage{booktabs}
\newcommand{\tabitem}{~~\llap{\textbullet}~~}
\def\BibTeX{{\rm B\kern-.05em{\sc i\kern-.025em b}\kern-.08em
    T\kern-.1667em\lower.7ex\hbox{E}\kern-.125emX}}

\usepackage{array}
\newcolumntype{?}{!{\vrule width 1pt}}

\newtheorem{remark}{Remark}

\else
 
\fi

\usepackage{nomencl}
\makenomenclature

\usepackage{etoolbox}
\renewcommand\nomgroup[1]{%
  \item[\bfseries
  \ifstrequal{#1}{I}{Sets and Indices}{%
  \ifstrequal{#1}{P}{Parameters}{%
  \ifstrequal{#1}{S}{Scenario Generation Variables}{%
  \ifstrequal{#1}{Q}{Decision Variables (Binary)}{%
  \ifstrequal{#1}{T}{Decision Variables}{}}}}}%
]}

\IEEEoverridecommandlockouts
\allowdisplaybreaks[2]

\begin{document}

\history{Date of publication xxxx 00, 0000, date of current version xxxx 00, 0000.}
\doi{}

\title{Two-stage Stochastic Optimal Power Flow for Microgrids with Uncertain Wildfire Effects}
\author{\uppercase{Sifat Chowdhury}, and
\uppercase{Yu Zhang}
\IEEEmembership{(Member, IEEE)}}

\address{Department of Electrical and Computer Engineering, University of California, Santa Cruz, CA 95064, USA}

\tfootnote{This work was supported in part by the Hellman Fellowship, in part by a Seed Fund Award from Center for Information Technology
Research in the Interest of Society (CITRIS) and the Banatao Institute, University of California, and in part by the UCSC CITRIS Interdisciplinary Innovation Program (I2P).}


\corresp{Corresponding author: Yu Zhang (e-mail: yzhan419@ucsc.edu).}

\begin{abstract}
Large-scale power outages caused by extreme weather events are one of the major factors weakening grid resilience. In order to prevent critical infrastructure from cascading failure, power lines are often proactively de-energized under the threat of a progressing wildfire. In this context, the potential of microgrid (MG) functioning in islanded mode can be exploited to enhance the resiliency of the power grid. However, there are numerous uncertainties originating from these types of events and an accurate modeling of the MG is required to harness its full potential. In this paper, we consider the uncertainty in line outages depending on fire propagation and reduced solar power generation due to the particulate matter in wildfire smoke. We formulate a two-stage stochastic MG optimal power flow problem by utilizing a second-order cone relaxation of the DistFlow model. Leveraging an effective approximation of the resistive heat gain, we separate the complicating constraints of dynamic line rating from the resulting optimization problem. Extensive simulation results corroborate the merits of our proposed framework, which is tested on a modified IEEE 22-bus system.
\end{abstract}

\titlepgskip=-21pt

\maketitle

\mbox{}

\nomenclature[I]{\(i,j/ij\)}{Indices for buses/Distribution line between i and j}
\nomenclature[I]{\(\mathcal{N}^{\text{MT}}/\mathcal{N}^{\text{PV}}/\mathcal{N}^{\text{WT}} \)}{Set of all buses connected with microturbines (MT) /solar PV / wind turbines (WT)}
\nomenclature[I]{\(\mathcal{N}^{\text{L}}/\mathcal{N}^{\text{QS}}\)}{Set of all load bus/ buses equipped with emergency generators or quick start (QS) units}
\nomenclature[I]{\(\mathcal{N}^{\text{ES}}/\mathcal{N}^{\text{MS}}\)}{Set of all buses having energy storage systems (ES) / mobile ES(MS)}
\nomenclature[I]{\({T^\prime}\)}{Look ahead window for first stage variables}
\nomenclature[I]{\(t\)/\(s\)}{Index of hours / Index for scenarios}
\nomenclature[I]{\(\mathcal{S}/\mathcal{T}\)}{Set of all scenarios / Set of hours for second stage variables $\mathcal{T}\triangleq \{1,2,\dots,T\}$}

\nomenclature[S]{\(\Phi^{\text{f}}\)}{Radiative heat flux emitted from wildfire (\si{W/m^{2}})}
\nomenclature[S]{\(\theta^{\text{f}}\)}{View angle between the flame and the conductor (\si{rad})} 
\nomenclature[S]{\(d^{\text{f}}\)}{Distance between fire and power line (\si{m})} 
\nomenclature[S]{\(q^{\text{s}}\)/\(q^{\text{f}}\)}{ Radiative heat gain rate from solar/fire (\si{W/m})} 
\nomenclature[S]{\(\check{T}/\vartheta^{\text{f}}\)}{Ambient temperature (\si{K}) / Fire spread rate(\si{m/sec})} 
\nomenclature[S]{\(u_{ij}\)}{Binary variable indicating line outage status}
\nomenclature[S]{\(q^{\text{l}}\)}{Resistive heat gain rate (\si{W/m})} 
\nomenclature[S]{\(q^{\text{r}}\)/\(q^{\text{c}}\)}{Radiative/Convective heat loss rate (\si{W/m})} 
\nomenclature[S]{\(T_{ij}\)}{Temperature of conductor connecting bus $i$ and $j$ (\si{K})}

\nomenclature[P]{\(\gamma^{\text{MT}}/\gamma^{\text{QS}}\)}{Marginal cost of active power generation from MT/QS units (\si{\$/MW})}
\nomenclature[P]{\(\gamma^{\text{MS}}\)}{Transportation cost of carrying an MS unit (\si{\$})}
\nomenclature[P]{\(\gamma^{\text{RF}}\)}{Price for buying per unit of renewable fuel (\si{\$/L})}
\nomenclature[P]{\(\pi_s\)}{Probability of each scenario}
\nomenclature[P]{\(\gamma^{\text{L}}\)}{Incentive for serving per unit of load (\si{\$/MW})}
\nomenclature[P]{\(\beta / \epsilon\)}{Load criticality factor / Solar absorptivity}
\nomenclature[P]{\(\tau^{\text{a}}\)}{Atmospheric transmissivity (dimensionless)}
\nomenclature[P]{\(\sigma^{\text{f}}/\sigma^{\text{c}}\)}{Emissivity of flame zone / conductor (dimensionless)}
\nomenclature[P]{\(\mu^{\text{a}}\)}{Thermal conductivity of air (\si{W/mK})}
\nomenclature[P]{\(\kappa^{\text{a}}\)}{Absolute viscosity of air (\si{kg/ms})}
\nomenclature[P]{\(C_{p}\)}{Relative heat capacity (\si{J/kg/K})}
\nomenclature[P]{\(\Phi^{\text{s}}/T_{\text{f}}\)}{Solar radiation (\si{W/m^{2}}) /Flame zone temperature (\si{K})}
\nomenclature[P]{\(\kappa_{\text{B}}\)}{Stefan–Boltzmann constant (\si{W/m^{2}/K^{4}})}
\nomenclature[P]{\(\varrho/ H\)}{Conductor diameter/Flame height (\si{m})}
\nomenclature[P]{\(F\)}{Effort level for extinguishing fire (dimensionless)}
\nomenclature[P]{\(M\)}{Indicator for uphill (0) and downhill (1)}
\nomenclature[P]{\(N\)}{Indicator for presence (10) or absence (0) of nearby natural obstacles}
\nomenclature[P]{\(\rho^{\text{a}}/\rho^{\text{b}}\)}{Air density / Density of bulk fuel (\si{kg/m^{3}})} 
\nomenclature[P]{\(\vartheta^{\text{w}}/\varphi\)}{Wind speed (\si{m/sec}) / Flame tilt angle (\si{rad})} 
\nomenclature[P]{\(\psi^{\text{w}}\)}{Angle between wind direction and conductor(\si{rad})}
\nomenclature[P]{\(r_{ij}/x_{ij}\)}{Line resistance/reactance between bus $i$ and bus $j$ (\si{\Omega})}
\nomenclature[P]{\(\zeta\)}{Thermal resistivity coefficient of the conductor (\si{K^{-1}})}
\nomenclature[P]{\(\bar{P}^{\text{MT}}\)}{MT maximum power output (\si{MW})}
\nomenclature[P]{\(k_{p}/k_{v}\)}{Droop constants}
\nomenclature[P]{\(\bar{P}^{\text{B}}/\bar{P}^{\text{S}}\)}{Maximum amount of power to be bought from/sold to the upstream network (\si{MW})}
\nomenclature[P]{\(\bar{P}^{\text{ES,ch}}/\bar{P}^{\text{ES,dis}}/\bar{P}^{\text{MS,ch}}/\bar{P}^{\text{MS,ch}}\)}{Upper limit of ES/ MS charging/discharging power (\si{MW})}
\nomenclature[P]{\(\eta^{\text{ch}}/\eta^{\text{dis}}\)}{ESS charging/discharging efficiency}
\nomenclature[P]{\(P^{\text{MT,ru}}/P^{\text{MT,rd}}\)}{MT ramping up/down limit (\si{MW})}
\nomenclature[P]{\(\eta^{\text{QS}}\)}{Conversion factor of QS fuel expenditure and power output}
\nomenclature[P]{\(\bar{Y}^{\delta}/\underline{Y}^{\delta}\)}{Upper/lower limit of QS fuel expenditure}
\nomenclature[P]{\(P^{\text{L}}/Q^{\text{L}}\)}{Active/reactive power demand (\si{MW/MVar})}
\nomenclature[P]{\(\bar{X}\)}{Maximum amount of fuel reserve that can be bought}
\nomenclature[P]{\(\bar{N}^{\text{MS}}/\bar{C}^{\text{MS}}\)}{Maximum number/allocated budget for MS units}
\nomenclature[P]{\(\bar{\ell_{ij}}\)}{Maximum squared magnitude of current flowing through line $ij$}

\nomenclature[T]{\(P_{i}/Q_{i}\)}{Active/reactive power injection at bus i (\si{MW/MVar})}
\nomenclature[T]{\(P_{ij}/Q_{ij}\)}{Active/reactive power flow through branch $ij$ (\si{MW/MVar})}
\nomenclature[T]{\(v_{i}\)}{Squared voltage magnitude at bus $i$}
\nomenclature[T]{\(\ell_{ij}\)}{Squared magnitude of current flowing through line $ij$}
\nomenclature[T]{\(P^{\text{ES,ch}}/P^{\text{ES,dis}}/P^{\text{MS,ch}}/P^{\text{MS,dis}}\)}{Scheduled ES/MS charging/ discharging active power (\si{MW})}
\nomenclature[T]{\(X_{i} / Y_i\)}{Renewable fuel reserve / Remaining fuel  of the $i$-th QS unit}
\nomenclature[T]{\(Z_{i}\)}{Binary indicator whether an MS unit is sited at bus $i$}
\nomenclature[T]{\(\alpha^{\text{ls}}/\alpha^{\text{ch}}/\alpha^{\text{dis}}\)}{Ratio of the served load to its actual demand/binary indicator for charging/discharging status for MS units}
\nomenclature[T]{\(S_i^{\text{S}} / S_i^{\text{M}}\)}{State of charge of the $i$-th ES/MS unit}

\printnomenclature

\section{Introduction}
\label{sec:introduction}
Extreme weather events pose a great threat to the system planning and operation of electrical grids. They severely affect the reliability and resiliency of the power system by damaging critical infrastructure like power lines and other electrical equipment. This leads to prolonged power outages to a large customer base and critical loads in the system. According to the U.S. Energy Information Administration, customers faced over eight hours of power interruption on average in 2020, which has doubled in just five years \cite{USEIA2021}. More frequent power outages due to high-impact, low-probability events result in tremendous financial losses, roughly \$150 billion per year \cite{USDOE2017}. As one of the most devastating natural disasters, wildfires take a toll on human beings and the environment by emitting tons of greenhouse gases. Moreover, to reduce the risk of power line induced fires, frequent public safety power shutoffs (PSPS) also cause huge economic losses. The disrupted power supply caused by a wildfire can be overcome by deploying microgrids (MG). A big advantage of an MG is that it can perform as an independent entity to serve its load when getting disconnected from the main grid. On top of that, MG makes the grid greener and more sustainable by integrating distributed energy resources (DER), including the microturbine (MT), renewable energy from photovoltaic (PV) panels, and wind turbines (WT).

\subsection{Motivation}

The optimal operation of an MG during a wildfire is subject to several uncertain factors. First of all, the progression of the fire and its degree of devastation depends upon a lot of features, including but not limited to wind speed, wind direction, and the amount of flammable fuel in the surrounding area. The onset of any contingency in the electric distribution system depends upon the fire dynamics. Furthermore, the generation portfolio of PV and WT is inherently stochastic, unlike conventional generators. During a wildfire, aerosols produced by the burning of organic substances visibly block sunlight from reaching the earth's surface and make the atmosphere darker. Therefore, the output from PV panel drops down due to the presence of smoke and soot containing different airborne particulate matter. According to California Independent System Operator (CAISO), the solar powered electricity generation declined more than 30\% in the first half of September 2020 in just 2 months when the state was experiencing one of the biggest fires in its history \cite{USEIA2021_2}. Similarly, the power output from WT and the load consumption profile of the customers are also contingent upon different scenarios. Therefore, these components need to be analyzed carefully to ensure the optimal scheduling of an MG.

\subsection{Literature Review}

Ensuring power supply to customers threatened by wildfire risks is a highly multi-disciplinary field of research encompassing geography, meteorology, engineering, and resource management. From a holistic point of view, we break down this broad topic into several key aspects as presented in Table \ref{Tab:lit review}. Forecasting an extreme weather event is the initial step in addressing the accompanying adverse impacts. To get off the ground, an effective model for wildfire prediction is essential to estimate the chance of a fire-induced power outage in a region. Authors from \cite{sifat_ctgan}, \cite{wfpred2}, \cite{wfpred3}, and \cite{wfpred4} investigate the underlying factors e.g. temperature, humidity, wind speed, vegetation condition, and topology that can initiate a fire. However, these works do not comprise a fire propagation model through which the risk and scale of an imminent fire can be determined. To overcome this challenge, \cite{trakas}, \cite{WF_dynamics_2} incorporate wildfire dynamics in their model for determining line outage by a progressing fire and achieving an optimal distribution system operation. A spatiotemporal fire monitoring technique in combination with a decision support tool for fire management is used by \cite{WF_dynamics_3} to mitigate the impacts of wildfires on the power grid. Dian et al. design a framework by integrating a prediction model and line outage probability to get an early warning of disrupted power lines \cite{WF_dynamics_4}. Nonetheless, their efforts are limited by strong assumptions, complicated problem formulation, and the absence of fire-extinguishing efforts.

Apart from the necessity of having dynamic wildfire prediction models, the energy management of an MG in a high fire risk zone is challenged by numerous uncertainties such as repair time of damaged equipment, islanded status, and intermittent renewable generation. \cite{uncertain_rt} co-optimizes system operation and repair crew routing considering the uncertainty in repair time and load demand. A stochastic optimization framework is developed for scheduling resources considering the uncertain duration of islanded operation of MG in \cite{islanding_duration_uncertainty}, \cite{islanding_uncertainty_2}. While their effort is important for the optimal operation of the power network, however, there are missing factors like fire extinguishing works that can alter the duration of the islanded condition. 

Renewable energy resources such as PV panels and WT are one of the key distributed generation components in MG which are inherently stochastic. The survey \cite{review} covers the recent advancements in stochastic optimization for modeling the uncertainties in renewable energy applications. The stochasticity of these resources is further captured through a robust optimization-based model in \cite{robust_res_uncertainty}, \cite{Applied_Energy5}, by an approximate dynamic programming-based approach in \cite{ADP} and as a security-constrained stochastic framework for the economic dispatch of flexible resources in \cite{Applied_Energy2}. Besides, several data-driven models are also proposed to approximate the uncertain sets originated by high penetration of renewable energy and load in \cite{load_and_wind}, using a hierarchical hybrid control method \cite{wang2018}, distributionally robust approach \cite{hybrid_res_uncertainty} and an adaptive robust formulation \cite{res_uncertainty_adaptive_robust}. Although these works significantly enhance the efficiency of renewable energy based models, they lack in assessing the adverse effects on solar power generation stemming from the fire smoke. Few works exist in the literature exploring the smoke effect by analyzing different molecules and particles produced by smoke \cite{smoke_effect}, \cite{Smoke_2}. \cite{smoke_4} develop a model to estimate the power output reduction on different PV cell technologies. Using the optical properties of smoke and the spectral response of PV cell materials, an estimated PV power reduction approach is presented in \cite{smoke_6}. Due to this sudden drop in power generation, \cite{smoke_5} studies the frequency stability on the power grid. Despite all, there remains a research gap in developing a compensation strategy for the unserved loads caused by the reduced generation.

Other diverse initiatives to make the power grid resilient against disasters are extensively studied in the literature. Public safety power shut-off is one of the emergency measures where electric components are selectively de-energized to reduce the risk of fire ignition. This task is formulated by \cite{Rhodes} and \cite{PSPS2} to minimize the risk of fire ignition from grid fault and maximize secure power delivery. Besides that, researchers also develop strategies for restoring loads during a natural disaster. \cite{lr_1} coordinates mobile resources, multiple generating units, and demand response, thus balancing the supply and demand side for restoring loads. Another collaborative strategy between network reconfiguration and spatio-temporal distribution of EV is explored in \cite{lr_2}  while \cite{DS_repair} simultaneously combines the repair process of the distribution system and load restoration problem. Network configuration is another way of serving critical loads affected by a natural disaster.
An advanced configuration of the distribution system is presented for restoring such loads \cite{Dubey}.

Besides, securing additional resources is another good practice that helps in overcoming the service disruption caused by an extreme weather event. From a resource allocation perspective, researchers propose a stochastic mixed integer linear programming model consisting of mobile and labor resources to assist load restoration in the post-event phase \cite{resource_allocation}. By forming adaptive multi microgrids with mobile energy resources \cite{Shahidehpour}, topology switching \cite{MG_formation}, resilience against extreme conditions is achieved. Besides, Liu et al formulate security constraints for spinning and non-spinning reserves in a limited energy resources setup against contingencies \cite{secured_NSR}. Moreover, there are efforts in advancing novel techniques to solve the MG energy management problem. A combination of two-stage stochastic programming and model predictive control is applied for MG energy management under uncertainties \cite{SP&MPC1},\cite{SP&MPC2}. A quantum teaching learning-based algorithm is also proposed by \cite{quantum} that captures the seasonal variation of solar and wind power generation.  Furthermore, demand response is incorporated in optimal energy dispatch strategies that potentially can decrease MG operation cost; see \cite{DR1}. One key shortcoming of these works is their failure to account for the uncertain effects of line outage caused by a progressing fire and smoke induced reduction in PV generation. Because of these effects, load service gets disrupted, which eventually affects the reliability and resilience of the power grid.

\begin{table*}[!t]
 \caption{Summary of the literature review}
 \label{Tab:lit review}
 \centering
 \begin{tabular}{l|l|l|l}
    \hline
    \textbf{Topic} & \textbf{References} & \textbf{Contribution} & \textbf{Limitation}\\  
    \hline
    Wildfire prediction & \cite{sifat_ctgan}, \cite{wfpred2},  & Analysis of meteorological factors & Lacks fire dynamics  \\ & \cite{wfpred3}, \cite{wfpred4} &causing wildfires & \\ 
    \hline
    Fire propagation model & \cite{trakas}, \cite{WF_dynamics_2}, & Determining line outage by tracking & \tabitem DLR constraints coupled to optimization problem
    \\
    & \cite{WF_dynamics_4}, \cite{WF_dynamics_3}  & fire progression & \tabitem Absence of human and natural intervention\\ & & & \tabitem Line outage being considered for tie-line only\\
    \hline
    Uncertainty modelling & & & \\
    \quad \tabitem Islanding duration  & \cite{islanding_duration_uncertainty}, \cite{islanding_uncertainty_2} & \tabitem The onset and duration of islanding mode & \tabitem Absence of fire extinguishing efforts \\ 
    &  & &\\
    \quad \tabitem Renewable energy & \cite{wang2018}, \cite{hybrid_res_uncertainty}, & \tabitem Addressing the inherent stochastic & \tabitem Absence of factors altering the\\ \quad generation& \cite{ADP}, \cite{Applied_Energy2}, & nature of energy  & predicted generation due to fire\\ & \cite{res_uncertainty_adaptive_robust}, \cite{robust_res_uncertainty}, & & \\ & \cite{Applied_Energy5} & & \\
    \hline
    Smoke effect & \cite{smoke_effect}, \cite{Smoke_2},  & Analysis of particulate materials present in smoke & Compensating the reduced solar power \\ & \cite{Smoke_3},\cite{smoke_4}, & & generation for serving loads\\ & \cite{smoke_5}, \cite{smoke_6} & & \\
    \hline
    Resilience initiatives & & & \\
    \quad \tabitem PSPS &\cite{Rhodes}, \cite{PSPS2} & \tabitem Reducing risks of fire ignition from power cables & \tabitem Load curtailment\\
    &  & &\\
    \quad \tabitem Load restoration & \cite{lr_1}, \cite{Dubey} & \tabitem Restoring curtailed loads & \tabitem Void of modelling and impacts of  \\
    &  & &disasters on the distribution network\\ & & &\\
    \quad \tabitem Resource allocation &  \cite{resource_allocation}, \cite{secured_NSR} & \tabitem Pre-event preparation for securing  & \tabitem Over estimation due to ignoring disaster-induced \\
    &  & additional resources &reduction of renewable energy generation\\ 
    \hline
\end{tabular}
\end{table*}

\subsection{Our Contributions}
In this work, we propose an MG scheduling task by formulating a two-stage
stochastic optimal power flow (OPF) problem considering two significant effects induced by an approaching wildfire, namely line outage and smoke-altered PV generation. We make reasonable assumptions to convexify the original non-convex problem and reduce computational complexity. 
The proposed framework is tested through extensive simulation to demonstrate the system’s resilience compared with some existing schemes. The novelty and contributions of our work are summarized as follows:

\begin{itemize}
    \item We comprehensively model wildfire propagation by incorporating natural and human exertion on top of meteorological and geographical factors. The inclusion of these factors captures both the progress and deflection of the fire front, which determines probable line outages.

    \item We model the curtailed PV generation originating from fire smoke using a data-driven approach. We assess the requirement of securing additional reserves to compensate for generation shortage. This ensures maximum load supply during service interruption. 

    \item We introduce a simple but effective approximation of the resistive heat gain rate, which facilitates the separation of the dynamic line rating constraints from the optimization problem. It helps in the convexification of the original problem and significantly reduces the computational burden.


\end{itemize}


The rest of this paper is organized as follows. Section II describes the system modeling consisting of wildfire dynamics and smoke effect. In Section III, the formulation of a two-stage stochastic optimization problem is discussed. Simulation studies are presented in Section IV. Finally, Section V provides the concluding remark along with the future research direction.

\section{System Modeling}
\subsection{Scenario Generation}
The uncertainties regarding the progression of wildfire and renewable generation are incorporated in our model as different scenarios. The stochastic parameters for these scenarios are wind speed ($\vartheta^{\text{w}}$), wind direction ($\psi^{\text{w}}$) and solar radiation ($\Phi^{\text{s}}$). They are modeled by the Weibull, von Mises and Beta distribution via historical data for each hour of the day, respectively. The other parameters considered are firefighting efforts, natural barriers, and regional topography as described in \eqref{Eqtn:fire spread rate}. Afterward, sufficient scenarios consisting of variables (listed earlier in the nomenclature section for scenario generation) as functions of these parameters are generated for 24 hours of a day using Monte Carlo simulation. The output power from the WT is calculated as mentioned in \cite{WT_op}. For the remaining parts of this paper, the subscripts $\{ij,s,t\}$ stand for the line connecting bus $i$ and $j$ at time $t$ for scenario $s$.

\subsection{Wildfire Modelling} \label{sec:WFM}
To incorporate the fire dynamics in our modeling, we leverage a classic heat flux model \cite{heat_flux_model}. This model overlooks human or natural interventions that could slow fire spread, as well as the influence of regional topography. To overcome this, our proposed fire propagation model is given as follows. 
\begin{subequations}
\label{eq:wildfire}
    \begin{align}
        \vartheta^{\text{f}}_{s,t} &=  0.07\times \frac{ (1+\vartheta^{\text{w}}_{s,t})}{\rho^{b}}  \times e^{\text{-F}} \times e^{\text{-N}} \times e^{-1^{\text{M}} \times \left|\text{slope} \right|} 
        \label{Eqtn:fire spread rate} \\
        d^{\text{f}}_{ij,s,t} &=  d^{\text{f}}_{ij,s,t-1}-\vartheta^{\text{f}}_{s,t}  \Delta t  \cos(\psi^{\text{w}}_{ij,s,t}) \label{Eqtn:fire distance calculation}\\
        \theta^{\text{f}} _{ij,s,t} &=  \arctan{\left(\frac{H \cos(\varphi )}{d^{\text{f}}_{ij,s,t} - H \sin(\varphi )}\right)} \label{Eqtn:view angle} \\
        \Phi^{\text{f}}_{ij,s,t} &= 0.5\times \sigma^{\text{f}}\kappa_{\text{B}} \tau^{\text{a}} T_{\text{f}}^{4} \sin(\theta^{\text{f}}_{ij,s,t}). \label{Eqtn:radiative heat}
    \end{align}
\end{subequations}
First, in \eqref{Eqtn:fire spread rate}, the fire spread rate $\vartheta^{\text{f}}$ is expressed as a function of wind speed $\vartheta^{\text{w}}$, bulk fuel density $\rho^{b}$, natural and human fire fighting efforts and the type of topography. Fuel density indicates the vegetation level of the area. We model the human exertion factor (firefighting effort) as an exponential function where $\text{F}$ quantifies different levels of effort. The rationale behind this modeling is that the growth or decay of any object is naturally modeled as an exponential function. $\text{F}$ can take three distinct values respectively, indicating no effort, moderate effort (use of fire hose from ground level), and extensive effort (combination of fire hose from ground and water planes from above. The next term ($\text{N}$) indicates the presence of natural barriers like water bodies and rocky outcrops that hinder fire growth. The last term ($\text{M}$) in this equation is the contributing factor by the landscape of an area \cite{WF_dynamics_4}. $\text{M}$ can be either 0 or 1, standing for uphill and downhill, respectively. Fire spreads faster on slanted land compared to flat land. In uphill terrain, radiant heat from a wildfire pre-heats fuels on the slope ahead of the fire front, facilitating faster propagation compared to downhill terrain. Afterward, \eqref{Eqtn:fire distance calculation} reflects the dynamics of the distance, $d^{\text{f}}$ between fire and the power line connecting bus $i$ and $j$ based on the wind speed and direction. Then, the view angle $\theta^{\text{f}}$ is calculated according to \eqref{Eqtn:view angle}, which depends on the distance $d^{\text{f}}$, flame height $H$ and flame tilt angle $\varphi $. Finally, we obtain the radiative heat flux $\Phi^{\text{f}}_{ij}$ emitted from the fire to the power line, as shown in \eqref{Eqtn:radiative heat}.

\subsection{Dynamic Line Rating (DLR)} \label{sec:DRL}
The temperature of an overhead conductor (power distribution line) is constantly changing in response to changes in weather conditions and the current flowing through it. In this subsection, we study the impact of the nearby fire and other factors on the conductor's temperature according to the IEEE standard 738-2012 \cite{IEEE_std_738}.

The dynamics of the line temperature $T_{ij,s,t}$ are quantified in heat balance equation~\eqref{Eqtn:HeatBal}, where three heat gain sources, respective from fire, solar and line current $\{ q^{\text{f}},q^{\text{s}},q^{\text{l}}\}_{ij,s,t}$ and two heat loss sources through radiation and convection $\{q^{\text{r}},q^{\text{c}}\}_{ij,s,t}$ are modeled by equations \eqref{Eqtn:fire heat radiation}-\eqref{Eqtn:convective heat loss}.
As shown in the \emph{big-M} constraint \eqref{Eqtn:line outage}, if $T_{ij}$ exceeds the maximum permissible temperature $\bar{T}_{ij}$, this line is switched off ($u_{ij}=0$) to ensure safe operation. 
\begin{subequations}
\label{eq:DRL}
    \begin{align}
        &T_{ij,s,t+1} - T_{ij,s,t} =  \nonumber \\ &\frac{\Delta t}{mC_{p}}\times  \big(q^{\text{f}}_{ij,s,t} + q^{\text{s}}_{ij,s,t}+ q^{\text{l}}_{ij,s,t} - q^{\text{r}}_{ij,s,t} - q^{\text{c}}_{ij,s,t} \big)
        \label{Eqtn:HeatBal}\\
        &T_{ij,s,t} \leq \bar{T}_{ij} + \left(1-u_{ij,s,t}\right)\times M
        \label{Eqtn:line outage} \\
        &q^{\text{f}}_{ij,s,t} = \varrho_{ij} \Phi^{\text{f}}_{ij,s,t} \label{Eqtn:fire heat radiation} \\
        &q^{\text{s}}_{ij,s,t} = \varrho_{ij} \epsilon_{ij} \Phi^{\text{s}}_{ij,s,t} \label{Eqtn:solar radiation} \\
        &q^{\text{l}}_{ij,s,t} = R_{ij,s,t}(T_{ij,s,t}) \times \ell_{ij,s,t}  \label{Eqtn:resistive heat gain} \\
        &q^{\text{r}}_{ij,s,t} = 1.78\times 10^{-7} \times \varrho_{ij} \times \sigma^{\text{c}}\times (T_{ij,s,t}^{4} - \check{T}_{ij,s,t}^{4}) \label{Eqtn:radiative heat loss}\\
        &q^{\text{c}}_{ij,s,t} =  \mu^{\text{a}} \chi_{ij,s,t}\times (T_{ij,s,t} - \check{T}_{ij,s,t}) \times f(\text{Re}_{ij,s,t}),
        \label{Eqtn:convective heat loss}
    \end{align}
\end{subequations}
where in \eqref{Eqtn:resistive heat gain} and \eqref{Eqtn:convective heat loss} we have 
\begin{align*}
R_{ij,s,t}(T_{ij,s,t}) &= R_{ij}^{\text{ref}} \times \left(1+ \zeta_{ij}\times (T_{ij,s,t} - T_{ij}^{\text{ref}})\right)\\
\text{Re}_{ij,s,t} &= \frac{\varrho_{ij}  \rho^{\text{a}} \vartheta^{\text{w}}_{s,t}}{ \kappa^{\text{a}}} \\
f(\text{Re}_{ij,s,t}) &= \max\left\{  1.01+1.35\times \text{Re}_{ij,s,t}^{0.52},\, 0.754 \times \text{Re}_{ij,s,t}^{0.6}\right\} \\
\chi_{ij,s,t} &= 1.194+0.194\times \cos(2\psi^{\text{w}}_{ij,s,t})\\& \hspace{1.3cm}+0.368 \times \sin(2\psi^{\text{w}}_{ij,s,t})- \cos(\psi^{\text{w}}_{ij,s,t}).
\end{align*}

\subsection{Smoke Effect}\label{smoke effect}
Solar panels produce energy by absorbing sunlight passing through the atmosphere. 
The total solar irradiance emitted from the sun is termed as global horizontal irradiance (GHI). However, the output from a solar panel depends upon a fraction of GHI that incidents perpendicularly on the solar panel, namely direct normal irradiance (DNI). The rest of the GHI is diffused by different molecules and particles in the atmosphere and contributes little to the PV generation.  Wildfire smoke can severely affect the PV generation capacity by blocking the effective solar irradiance. The presence of particulate matter (PM), specifically PM2.5 and PM10 in the smoke plume mainly affects PV production while clouds and some other meteorological parameters contribute there too. The atmospheric clouds reflect and scatter the sunlight and prevent them from reaching the PV panels. Hence, the thicker the cloud is, the less the power generation will be. Similarly, the higher concentration of PM  attenuates PV output. Moreover, a higher relative humidity indicates the presence of more water vapor in the atmosphere, which causes more refraction and reflection of direct solar insolation.

In \cite{Smoke_3}, the effect of a short-span prescribed fire burn on solar irradiance and PV power was investigated. There was no conclusive correlation among the factors causing the PV output reduction. \cite{Smoke_2} studied the effect of aerosol on surface solar radiation considering a clear sky condition. \cite{trakas} used Beta distribution to model the GHI while ignoring crucial factors such as the existence of airborne PM caused by a fire. Without considering the cloud and all types of PM, \cite{smoke_effect} measured the effect of certain environmental factors causing a reduced PV generation. Hence, during a contingency, the PV generation forecast can be highly overestimated. Moreover, a theoretical model was developed by \cite{smoke_6} to estimate the PV output reduction by using the optical properties of smoke. The impact on frequency stability arising from this sudden reduction in generation on the power grid was analyzed by \cite{smoke_5}. \cite{smoke_4} developed another data-driven model to study how solar power reduction varies in different PV cell technologies.

To this end, in order to quantify the smoke effect more accurately, we postulate the following multivariate regression model:
\begin{align}
P^{\text{PV}}_{s,t} = \alpha_{0} + \alpha_{1}Z_{1} + \alpha_{2}Z_{2} &+ \alpha_3  Z_{3} +  \alpha_4 Z_{4} \notag \\ & +\alpha_5\ln(Z_{5}) + \alpha_6\ln(Z_{6}),
\label{Eqtn:smoke_effect}
\end{align}
where the target value is the measured PV generation output $P^{\text{PV}}$. The features include temperature $Z_{1}$, relative humidity $Z_{2}$, DNI on the PV panel $Z_{3}$, cloud opacity $Z_{4}$, the concentrations of PM2.5 $Z_{5}$ and PM10 $Z_{6}$. For the last two factors, natural logarithm is applied because the solar irradiation changes exponentially with the presence of the particulate matter. $Z_{3}$ is the net solar irradiation reaching the PV panels, which is equal to the DNI multiplied by the cosine of solar zenith angle $\theta_{z}$. All the variables are standardized by subtracting their mean values and divided by the standard deviations.

\section{Problem Formulation}
We represent an MG as a directed tree graph $\mathcal{G}\triangleq (\mathcal{N},\mathcal{E})$, where $\mathcal{N}$ is the set of buses and $\mathcal{E}$ denotes the set of power lines (branches). We assume that buses can either be connected with a load, microturbines, quick start generators, wind turbines or PV panels. In addition, all PV and WT buses are equipped with energy storage systems and a mobile energy storage unit can be sited to any bus other than the load buses. Therefore, we have $\mathcal{N} = \mathcal{N}^{\text{L}} \cup \mathcal{N}^{\text{MT}} \cup \mathcal{N}^{\text{QS}} \cup \mathcal{N}^{\text{PV}} \cup \mathcal{N}^{\text{WT}}$, $\mathcal{N}^{\text{ES}}= \mathcal{N}^{\text{PV}} \cup \mathcal{N}^{\text{WT}}$ and $\mathcal{N}^{\text{MS}}= \mathcal{N} \setminus  \mathcal{N}^{\text{L}}$. Unless specified otherwise, the objective function and constraints described in this section hold for all scenarios $s \in \mathcal{S}$.

\subsection{Objective Function}\label{sec:ObjFun}
Considering the inherent uncertainty of renewable energy generation and wildfire dynamics, we formulate the task of MG optimal power scheduling as a two-stage stochastic optimization problem. The objective function \eqref{Eqtn:obj} minimizes the costs of power generation and load curtailment that occurred in the first and second stages. The total time horizon is divided into two segments, $T^\prime$ and $\mathcal{T}$, which respectively realize the first and second stage variables. The first stage variables include siting an MS unit $\{Z_i\}_{i\in \mathcal{N}\setminus \mathcal{N}^{\text{L}}}$ to bus $i$ and the allocated fuel reserves $\{X_i\}_{i\in \mathcal{N}^{\text{QS}}}$ for the QS generators. 
Then, we generate a sufficiently large number of scenarios consisting of various fire dynamics, weather conditions and the smoke effect. In the second stage, the goal is to minimize the total expected operation cost while maximizing prioritized load serving. The variables include real-time dispatch commands and other network operational variables in the MG.
\begin{equation} 
{\text{minimize}}\,\,  C_1\left(\{Z_{i},X_{i}\}_i\right) + \color{black}C_2\left(\{P_{i,s,t},\alpha^{\text{ls}}_{i,s,t}\}_{i,s,t}\right) \\
\label{Eqtn:obj}
\end{equation}
where 
\begin{subequations}
\begin{align}
& C_1\left(\{Z_{i},X_{i}\}_i\right) = \sum_{i \in \mathcal{N} \setminus \mathcal{N}^{\text{L}} } \gamma^{\text{MS}} Z_{i}  + \sum_{i \in \mathcal{N}^{\text{QS}}}  \gamma^{\text{RF}} X_{i}
\label{Eqtn:c1}\\
&C_2\left(\{P_{i,s,t},\alpha^{\text{ls}}_{i,s,t}\}_{i,s,t}\right) = \sum_{s \in \mathcal{S}}\sum_{t \in \mathcal{T}} \pi_{s} \times \Bigg( \sum_{i \in \mathcal{N}^{\text{QS}}}  \gamma^{\text{QS}}P_{i,s,t}  \notag\\ & \hspace{1cm} + \sum_{i \in \mathcal{N}^{\text{MT}}}  \gamma^{\text{MT}}P_{i,s,t}  - \sum_{i \in \mathcal {N}^\text{{L}}}\gamma^{\text{L}}\alpha^{\text{ls}}_{i,s,t} \beta_{i,t} P^{\text{L}}_{i,s,t} \Bigg)
\label{Eqtn:c2}
\end{align}
\end{subequations}
The term $\alpha^{\text{ls}}_{i,t,s} \beta_{i,t} P^{\text{L}}_{i,t}$ reflects the prioritized load serving amount. In order to prioritize demand satisfaction, the incentive for serving loads is set large enough.

\subsection{First Stage Constraints}

The variables in this stage are decided $T^\prime$ hours before the second stage scheduling begins. The siting of MS units is constrained by the number of units and the potential budget allocated for them. \eqref{number of MS unit} and \eqref{budget for MS} denote these constraints. For simplicity, we assume that all the units are of the same size, fully charged and each bus can accommodate only one MS unit. \eqref{buying RF for each unit} limits the summation of the total amount of renewable fuel reserve that can be allocated to all QS units, $i \in \mathcal{N}^{\text{QS}}$. 
\begin{subequations}
\label{cstr:1stage}
\begin{align}
& \sum_{i \in \mathcal{N} \setminus \mathcal{N^{\text{L}}}}  Z_{i} \leq \bar{N}^{\text{MS}}
\label{number of MS unit}\\
& \sum_{i \in \mathcal{N} \setminus \mathcal{N^{\text{L}}}}  \gamma^{\text{MS}} Z_{i} \leq \bar{C}^{\text{MS}}
\label{budget for MS}\\
&0\le \sum_{i \in \mathcal{N}^{\text{QS}}} X_{i}  \leq \bar{X}. 
\label{buying RF for each unit}
\end{align}
\end{subequations}

\subsection{Second Stage Constraints}
This subsection explains the operational constraints of the MG for all time $t \in \mathcal{T}$. 

\subsubsection{Power Flow Constraints}
 For $j \in \mathcal{N}$ and $(i,j) \in \mathcal{E}$, an SOCP relaxation of the \emph{DistFlow model} \eqref{pf_1}-\eqref{pf_4} is adopted to quantify the power flows in the radial network; see e.g., \cite{Distflow}. Furthermore, \eqref{voltage_limit}-\eqref{current_limit} set the lower and upper limits for the nodal voltages and branch currents to ensure safe operation. The voltage of the reference bus is also fixed at 1 p.u. 
\begin{subequations}
\begin{align}
&P_{j,s,t} = \sum_{k:j\,\to\,k} P_{jk,s,t} - \sum_{i:i\,\to\,j} \left(P_{ij,s,t} - r_{ij}\ell_{ij,s,t} \right)
\label{pf_1}\\
&Q_{j,s,t} = \sum_{k:j\,\to\,k} Q_{jk,s,t} - \sum_{i:i\,\to\,j} \left(Q_{ij,s,t} - x_{ij}\ell_{ij,s,t} \right)
\label{pf_2}\\
&v_{j,s,t} = v_{i,s,t} - 2 \left(r_{ij}P_{ij,s,t} + x_{ij}Q_{ij,s,t}\right) + (r_{ij}^{2}+x_{ij}^{2})\ell_{ij,s,t}
\label{pf_3}\\
&P_{ij,s,t}^{2} + Q_{ij,s,t}^{2}  \le \ell_{ij,s,t} v_{j,s,t} \label{pf_4}\\
&\underline{v}\leq v_{j,s,t}\leq \bar{v}, \quad v_{s,t}^{\text{ref}}=1.
\label{voltage_limit}\\
&0\le \ell_{ij,s,t}\leq \bar{\ell}_{ij}.
\label{current_limit}
\end{align}
\end{subequations}

\subsubsection{Reactive Power Constraints}
We assume that all the buses except for the load bus are equipped with controllable inverters. Hence,  reactive power constraints are given as
\begin{align}
|Q_{i,s,t}| \leq \bar{Q}_{i},\quad  \forall i \in  \mathcal{N}\setminus \mathcal{N}^{\text{L}}.
\label{reactive power}
\end{align}

\subsubsection{QS Constraints}
To be environmentally friendly and cost effective, the MG leverages renewable fuels for quick start (QS) units as non-spinning reserves to restore the loads after an islanding event with PSPS.
The following constraints hold for all $i \in \mathcal{N}^{\text{QS}}$. $Y_{i,s,t}$ is the remaining fuel in the $i$-th QS unit. \eqref{initial fuel level} sets the initial fuel remaining in the $i$-th unit as the amount of fuel allocated to that unit in the first stage. This is a coupling constraint between the first and second stages. \eqref{ramping up} ensures that these units only start to operate at the next time step after any of the lines gets tripped off together with a limit for ramping up and down of the fuel expenditure \eqref{power output from QS units} relates the fuel expenditure with the amount of power produced with a conversion factor $\eta^{\text{QS}}$. \eqref{nonnegative fuel level} represents the non-negativity constraints for the remaining fuel level and power output, respectively.
\begin{subequations}
\begin{align}
&Y_{i,s}^{\text{int}} = X_{i}
\label{initial fuel level}\\
&  \prod_{ij \in \mathcal{E}}(1-u_{ij,s,t})\times \underline{Y}_{i}^{\delta} \leq Y_{i,s,t} - Y_{i,s,t+1} \notag\\ &  \quad \leq \prod_{ij \in \mathcal{E}}(1-u_{ij,s,t})\times \bar{Y}_{i}^{\delta},  \qquad \forall t \in  \mathcal{T} \setminus \{T\}
\label{ramping up}\\
&P_{i,s,t+1} = \eta^{\text{QS}} \times ( Y_{i,s,t} - Y_{i,s,t+1}), \quad \forall t \in  \mathcal{T} \setminus \{T\} 
\label{power output from QS units}\\
& Y_{i,s,t} \ge 0, P_{i,s,t} \ge 0.
\label{nonnegative fuel level}
\end{align}
\end{subequations}

\subsubsection{MT Constraints}
The generation limits and ramping up/down constraints for each MT unit at node $i \in \mathcal{N}^{\text{MT}}$ are given as follows.
\begin{subequations}
\begin{align}
&0 \leq P_{i,s,t} \leq \bar{P}^{\text{MT}}_{i}
\label{MT limit} \\
&P^{\text{MT,rd}}_{i} \leq P_{i,s,t+1} - P_{i,s,t}\leq P^{\text{MT,ru}}_{i},\, \forall t \in  \mathcal{T} \setminus \{T\}
\label{ramping up-down limit}
\end{align}
\end{subequations}

\subsubsection{Load Serving Constraints}
For each load at node $i \in \mathcal{N}^{\text{L}}$, the variable $\alpha^{\text{ls}}$ is the ratio of the served load to the actual load demand. With a constant power factor at all load buses, the condition is expressed as
\begin{align}
0 \leq \alpha_{i,s,t}^{\text{ls}} = &\frac{P_{i,s,t}}{P_{i,s,t}^{\text{L}}} = \frac{Q_{i,s,t}}{Q_{i,s,t}^{\text{L}}} \leq 1\, .
\label{eq:alphaLS} 
\end{align}

\subsubsection{Static Energy Storage Constraints}
To have fast responses with high ramping rates, static energy storage systems or ES units are used as an alternative to traditional spinning reserves. These systems come into effect immediately after a power outage.
For every ES unit $i \in  \mathcal{N}^{\text{ES}}$, we have the following constraints. \eqref{SOC1} is the dynamic equation of the state of charge (SoC) with the amounts of charging and discharging power $\{P^{\text{ES,ch}}, P^{\text{ES,dis}}\}$, which are associated with the efficiency factors $\{\eta^{\text{ch}}, \eta^{\text{dis}}\}$. The initial SoC, and lower/upper limits for all the ES units are specified in \eqref{SOC_initial}. \eqref{SOC2} ensures the charging and discharging status with the appropriate upper limit. To prevent simultaneous charging and discharging, a complementarity constraint \eqref{SOC3} is enforced. Finally, \eqref{SOC4} gives the net real power injection at any ES bus.

\begin{subequations}
\label{cstr:ES}
\begin{align}
&S_{i,s,t+1}^{\text{S}} = S_{i,s,t}^{\text{S}} + \eta^{\text{ch}} \Delta t P^{\text{ES,ch}}_{i,s,t} - \frac{1}{\eta^{\text{dis}}}\Delta t P^{\text{ES,dis}}_{i,s,t},\, \notag\\ &\hspace{150pt} \forall t \in  \mathcal{T} \setminus \{T\}
\label{SOC1}   \\
&S_{i,s,1}^{\text{S}} = S_{i,s}^{\text{S,int}}, \quad \underline{S} \leq S_{i,s,t}^{\text{S,int}} \leq \bar{S} 
\label{SOC_initial}   \\
&0 \leq P_{i,s,t}^{\text{ES,ch}} \leq \bar{P}^{\text{ES,ch}}_{i}, 0\leq P_{i,s,t}^{\text{ES,dis}} \leq \bar{P}^{\text{ES,dis}}_{i}
\label{SOC2}  \\ 
&P_{i,s,t}^{\text{ES,ch}} \times {P}^{\text{ES,dis}}_{i,s,t} = 0
\label{SOC3} \\
&P_{i,s,t} = P_{i,s,t}^{\text{ES,dis}} - P_{i,s,t}^{\text{ES,ch}}.
\label{SOC4}  
\end{align}
\end{subequations}

\subsubsection {Mobile Energy Storage Constraints}
The mobile storage or MS units follow the same operating principle as ES.  \eqref{MSOC1}-\eqref{MSOC2} denote the SoC constraints and net real power injections. Coupling constraints \eqref{MSOC3} and \eqref{MSOC4} denote the charging and discharging power limit where an MS unit is sited. These units can either be charged, discharged or remain idle at any time as specified by the binary indicators $\alpha^{\text{ch}}$ and $\alpha^{\text{dis}}$ in \eqref{MSOC5}. 
\begin{subequations}
\label{cstr:MS}
\begin{align}
&S_{i,s,t+1}^{\text{M}} = S_{i,s,t}^{\text{M}} + \eta^{\text{ch}} \Delta t P^{\text{MS,ch}}_{i,s,t} - \frac{1}{\eta^{\text{dis}}}\Delta t P^{\text{MS,dis}}_{i,s,t},\, \notag\\ &\hspace{150pt} \forall t \in  \mathcal{T} \setminus \{T\}
\label{MSOC1}   \\
&S_{i,s,1}^{\text{M}} = S_{i,s}^{\text{M,int}}, \quad \underline{S} \leq S_{i,s,t}^{\text{M}} \leq \bar{S} 
\label{MSOC_initial}   \\
& P_{i,s,t} = P_{i,s,t}^{\text{MS,dis}} - P_{i,s,t}^{\text{MS,ch}}.
\label{MSOC2}  \\
&0 \leq P_{i,s,t}^{\text{MS,ch}} \leq Z_{i} \times \alpha^{\text{ch}}_{i,s,t} \times \bar{P}^{\text{MS,ch}}_{i}
\label{MSOC3}  \\ 
&0\leq P_{i,s,t}^{\text{MS,dis}} \leq Z_{i} \times \alpha^{\text{dis}}_{i,s,t} \times \bar{P}^{\text{MS,dis}}_{i}
\label{MSOC4} \\
& \alpha^{\text{ch}}_{i,s,t} +  \alpha^{\text{dis}}_{i,s,t} \leq 1.
\label{MSOC5}
\end{align}
\end{subequations}

\subsection{Optimization Problem and Convexification}
To this end, we formulate the following stochastic OPF problem for the MG facing a progressing wildfire. 
\begin{mini}|l|
{\mathcal{V}_{1},\mathcal{V}_{2},\mathcal{V}^{\text{DLR}}}
{C_1\left(\mathcal{V}_{1}\right)+C_2\left(\mathcal{V}_{2}\right)}
{}{}
\addConstraint{\eqref{eq:DRL},\, \eqref{cstr:1stage}-\eqref{cstr:MS}.}
\label{OptFinal}
\end{mini}
The optimization variables 
$\mathcal{V}_{1} \triangleq \{X_{i},Z_{j}\}_{i \in \mathcal{N^{\text{QS}}}, j \in \mathcal{N} \setminus\mathcal{N^{\text{L}}}, t\in \mathcal{T}}$ and 
$\mathcal{V}_{2} \triangleq \{\mathcal{V}^{\text{MT}}, \mathcal{V}^{\text{QS}},\mathcal{V}^{\text{B}},\mathcal{V}^{\text{NC}},\mathcal{V}^{\text{ES}}, \mathcal{V}^{\text{MS}},\mathcal{V}^{\text{L}},\mathcal{V}^{\text{DLR}}\}$ are defined as follows.
\begin{align*}
\mathcal{V}^{\text{MT}} &\triangleq \{ P_{i,s,t}, Q_{i,s,t}\}_{i \in \mathcal{N^{\text{MT}}}, s\in \mathcal{S},t\in \mathcal{T}}\\
\mathcal{V}^{\text{QS}} &\triangleq \{ P_{i,s,t}, Q_{i,s,t},Y_{i,s,t}\}_{i \in \mathcal{N^{\text{QS}}}, s\in \mathcal{S},t\in \mathcal{T}}\\
\mathcal{V}^{\text{B}} &\triangleq \{\ell_{ij,s,t}, P_{ij,s,t},Q_{ij,s,t}\}_{(i,j) \in \mathcal{E}, s\in \mathcal{S},t\in \mathcal{T}}\\
\mathcal{V}^{\text{NC}} &\triangleq \{v_{i,s,t}\}_{i \in \mathcal{N}, s\in \mathcal{S},t\in \mathcal{T}}\\
\mathcal{V}^{\text{ES}} &\triangleq \{P_{i,s,t}^{\text{ES,ch}},P_{i,s,t}^{\text{ES,dis}},S_{i,s,t}^{\text{S}},P_{i,s,t}\}_{i \in \mathcal{N^{\text{ES}}}, s\in \mathcal{S},t\in \mathcal{T}}\\
\mathcal{V}^{\text{MS}} &\triangleq \{P_{i,s,t}^{\text{MS,ch}},P_{i,s,t}^{\text{MS,dis}},S_{i,s,t}^{\text{M}},P_{i,s,t},\alpha^{\text{ch}}_{i,s,t},\alpha^{\text{dis}}_{i,s,t}\}_{i \in \mathcal{N^{\text{MS}}}, s\in \mathcal{S},t\in \mathcal{T}}\\
\mathcal{V}^{\text{L}} &\triangleq \{\alpha_{i,s,t}^{\text{ls}}\}_{i \in \mathcal{N^{\text{L}}}, s\in \mathcal{S},t\in \mathcal{T}} \\
\mathcal{V}^{\text{DLR}} &\triangleq \{T_{ij,s,t},u_{ij,s,t},\ell_{ij,s,t}\}_{s\in \mathcal{S},t\in \mathcal{T}}.
\end{align*}

It can be seen that the resulting problem \eqref{OptFinal} is highly nonconvex due to \eqref{Eqtn:resistive heat gain}, \eqref{Eqtn:radiative heat loss}, \eqref{pf_4}, \eqref{ramping up}, \eqref{SOC3}, \eqref{MSOC3} and \eqref{MSOC4}. Among them \eqref{pf_4} is relaxed to second order cone constraints \cite{convex_hull}. \eqref {SOC2}-\eqref{SOC4} is replaced with the convex hull of its non-convex feasible region whose closed form is given in \cite{convex_hull}.
\begin{subequations}
\begin{align} 
    &0\leq P_{i,s,t}^{\text{ES,ch}} \\
    &0\leq P_{i,s,t}^{\text{ES,dis}}\\
    &\frac{P_{i,s,t}^{\text{ES,ch}}}{\bar{P}^{\text{ES,ch}}_{i}} + \frac{P_{i,s,t}^{\text{ES,dis}}}{\bar{P}^{\text{ES,dis}}_{i}} \leq 1. 
\end{align}
\end{subequations} 

We have bilinear terms in \eqref{MSOC3} and \eqref{MSOC4}. To convexify, we introduce two new sets of binary variables $\tau^{\text{ch}}$ and $\tau^{\text{dis}}$ and replace the said equations as follows. 
\begin{subequations}
\begin{align} 
    &\tau^{\text{ch}}_{i,s,t}\leq Z_{i}, \quad \tau^{\text{dis}}_{i,s,t} \leq Z_{i}  \\
    &\tau^{\text{ch}}_{i,s,t}\leq \alpha^{\text{ch}}_{i,s,t}, \quad \tau^{\text{dis}}_{i,s,t} \leq \alpha^{\text{dis}}_{i,s,t}  \\
    &\tau^{\text{ch}}_{i,s,t} \geq \left(Z_{i}+\alpha^{\text{ch}}_{i,s,t} -1 \right), \tau^{\text{dis}}_{i,s,t} \geq \left(Z_{i}+\alpha^{\text{dis}}_{i,s,t} -1 \right)\\
    &0\leq P_{i,s,t}^{\text{MS,ch}} \leq \tau^{\text{ch}}_{i,s,t} \times \bar{P}^{\text{MS,ch}}_{i}\\
    &0\leq P_{i,s,t}^{\text{MS,dis}} \leq \tau^{\text{dis}}_{i,s,t} \times \bar{P}^{\text{MS,dis}}_{i}.
\end{align}
\end{subequations}

For handling the non-convexity in \eqref{Eqtn:resistive heat gain} and \eqref{Eqtn:radiative heat loss}, \cite{trakas} uses a simple linearization technique as shown in \eqref{Eqtn:linearized resistive heat gain}-\eqref{Eqtn:linearized radiative heat loss}. Specifically, by fixing the line temperature at its maximum allowable value $\bar{T}^{ij}$, the outer approximation \eqref{Eqtn:linearized resistive heat gain} acts as a surrogate for the resistive heat gain \eqref{Eqtn:resistive heat gain}.  
Regarding the radiative heat loss \eqref{Eqtn:radiative heat loss}, 
it is linearized with respect to the line temperature as shown in \eqref{Eqtn:linearized radiative heat loss}. 
\begin{subequations}
\begin{align}
&q^{\text{l}}_{ij,s,t} \geq R_{ij,s,t} (\bar{T}^{ij}) \times \ell{ij,s,t} \label{Eqtn:linearized resistive heat gain}\\
&q^{\text{r}}_{ij,s,t} = \lambda^{\text{r}} \times T_{ij,s,t} + \beta^{\text{r}}. \label{Eqtn:linearized radiative heat loss}
\end{align}
\end{subequations}
By using these approximations, \eqref{eq:DRL} becomes a set of linear equality and inequality constraints in $\mathcal{V}^{\text{DLR}}$, which results in a mixed-integer second-order cone program.  

In this work, we observe the fact that the resistive heat gain $q^{\text{l}}_{ij,s,t}$ contributes much less to the line temperature change, compared with all other heat sources in \eqref{Eqtn:HeatBal}. Therefore, when dealing with the bilinear term $\ell_{ij,s,t}\times T_{ij,s,t}$ in \eqref{Eqtn:resistive heat gain}, we fix the line current at its upper limit. Consequently, the resistive heat gain $q^{\text{l}}_{ij,s,t}$  becomes a linear function in line temperature only, as given below. 
\begin{align}
q^{\text{l}}_{ij,s,t} = R_{ij,s,t} (T_{ij,s,t}) \times \bar{\ell}^{ij}.
\label{cnst:DLRdetach}
\end{align}

\begin{remark}
The advantage of our proposed approximation is threefold. First, separating the DLR constraints aids in reducing the number of decision variables by removing the squared magnitude of current ($\ell_{ij,s,t}$) and the line outage indicator ($u_{ij,s,t}$) from the optimization problem. As $u_{ij,s,t}$ can be obtained by solving \eqref{eq:DRL} independently, the computational time gets reduced to one-tenth according to our simulation. Therefore, we can afford to include more practical constraints in the problem. Second, separating DLR also eliminates the line temperature ($T_{ij}$) from being a decision variable.
Hence, we avoid the need to make strong assumptions to linearize the quartic function in equation (2f), which distinguishes our approach from certain traditional methods  \cite{trakas}, \cite{WF_dynamics_2}. By solving it outside the main optimization problem, we can preserve its integrity without adding complexity, thus avoiding unnecessary approximation errors. Third, removing $u_{ij,s,t}$ as a decision variable inherently resolves the non-convexity issue in \eqref{ramping up}.
\end{remark}

\begin{figure}[!tb]
	\centering
	\includegraphics[width=1.0\linewidth]{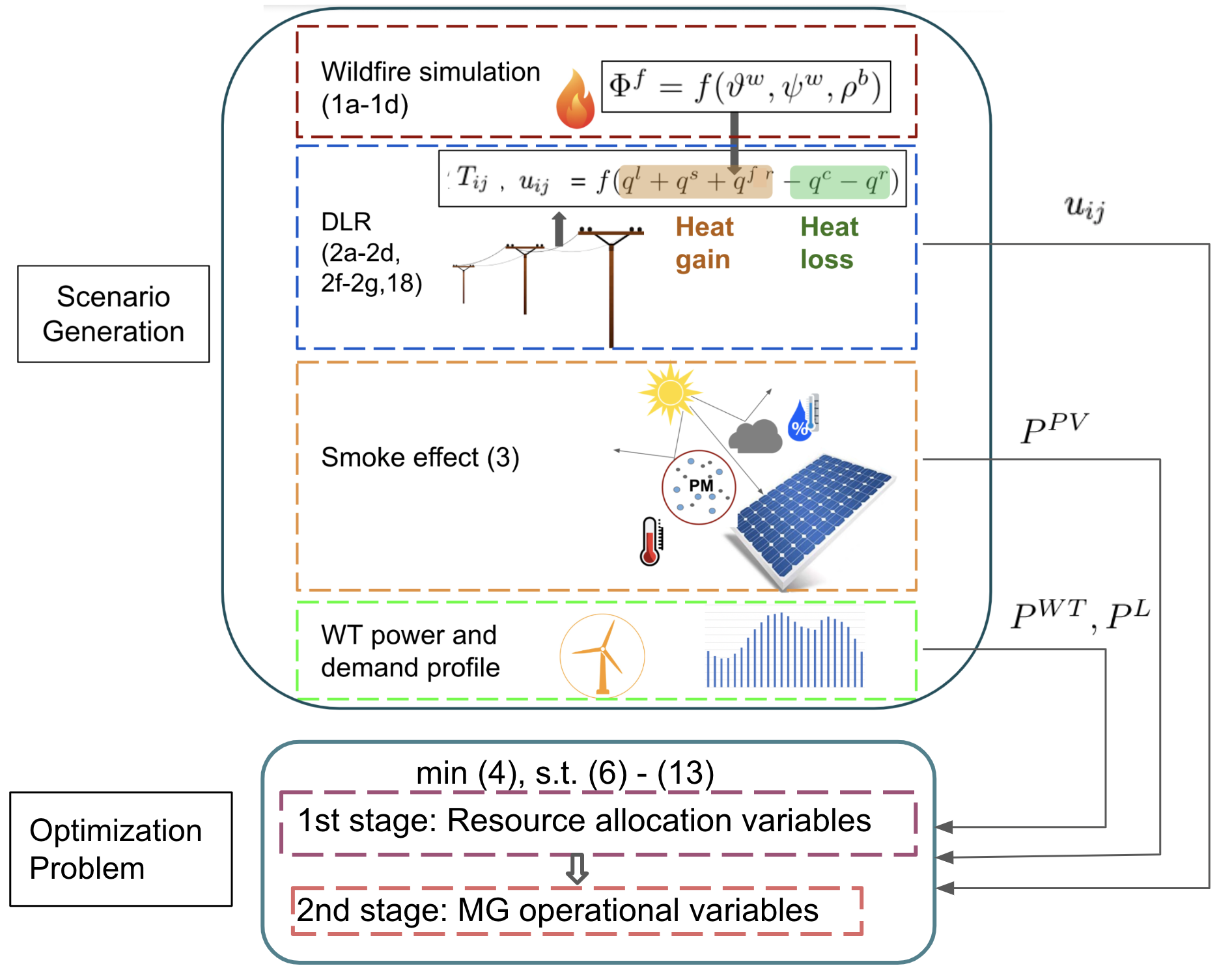}
	\caption{Framework of our proposed model: the generated renewable power, load demand and tie line outage status serve as the inputs of the resulting optimization problem from which the DLR constraints are separated.}
	\label{fig:framework}
\end{figure}

Finally, we illustrate the overall scheme of the proposed framework in  Figure~\ref{fig:framework}. Derived from the scenario generation, binary indicator $u_{ij}$, renewable power generation $P^{\text{PV}}, P^{\text{WT}}$, as well as the load demand $P^{\text{L}}$ are provided as the optimization problem inputs. 

\section{Simulation Results}

In this section, we present extensive simulation results to show the merits of our proposed framework. 

\subsection{Simulation Setup}
\begin{figure}[t!]
	\centering
	\includegraphics[scale=0.3]{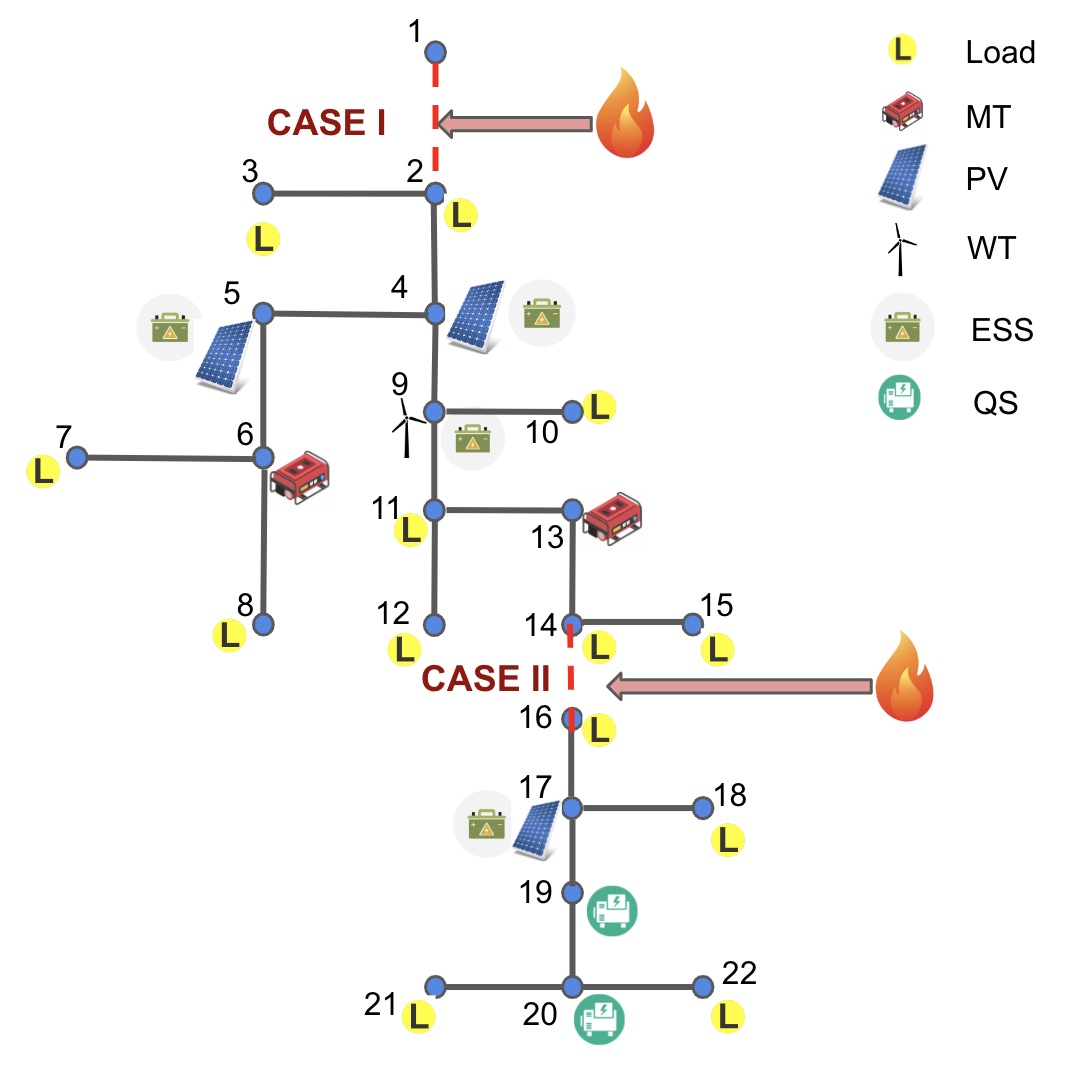}
	\caption{Outage lines and DER in the simulated system.}
	\label{fig:outage_lines}
\end{figure}

All the simulations are carried out by a PC with Intel Core i7, 3.6 GHz CPU and 32GB RAM. A modified IEEE 22-bus distribution system is considered, which has 3 PV units, 1 WT, 2 MT, and 2 QS generators. An ES is installed with each of the renewable units. We consider two different cases of line outages. Case I represents a line outage between buses 1 and 2 while case II is between buses 14 and 15 as specified in figure \ref{fig:outage_lines}. The network parameters are listed in Table \ref{Tab:Network parameters}. The demand profile data are collected from \cite{load_data}. To focus on the smoke effect on PV generation, we run the simulations for daytime from 7 am until 9 pm. We generate 50 scenarios and use one of them to show the scheduling results by the figures in Section \ref{MG_operation}. We use Gurobi 9.3 \cite{gurobi} along with CVX \cite{cvx} to solve the resulting SOCP. 

\begin{table} [!tb]
 \caption{Values of the Network Parameters}
 \label{Tab:Network parameters}
 \centering
 \begin{tabular}{c|c}
    \hline
    \textbf{Parameter} & \qquad \textbf{Value}\\  
    \hline
    $\mathcal{N}^{\text{L}}$ & \{2,3,7,8,10,11,12,14,15,16,18,21,22\}\\
    \hline
    $\beta$  & 2, 2, 3, 4, 8, 3, 3, 8, 9, 2, 11, 12, 7\\
    \hline 
    T, $T^\prime$ $\Delta$t & 15, 4, 1hr\\
    \hline
    $\gamma^{\text{B}},\gamma^{\text{S}},\gamma^{\text{RF}},\gamma^{\text{QS}},\gamma^{\text{MT}},\gamma^{\text{L}}$ &50, 50, 200, 65, 70, 10000 \\
    \hline
    $\underline{v},\bar{v}$ & 0.95, 1.05\\
    \hline
    $\bar{P}^{\text{MT}}, P^{\text{MT,ru}}, P^{\text{MT,rd}}$ & 0.05, 0.03, -0.02\\
    \hline
    $\bar{Y}^{\delta},\underline{Y}^{\delta}, \eta^{\text{QS}}$ & \{0.06, 0.07\}, \{0.025, 0.03\}, 0.6\\
    \hline $\eta^{\text{dis}},\eta^{\text{ch}},S^{\text{int}}$ & 0.9, 0.9, 0.4\\
    \hline 
    $\bar{P}^{\text{ES,dis}}$ & \{0.0328, 0.03, 0.0315, 0.0290\}\\
    \hline
    $\bar{P}^{\text{ES,ch}}$ & \{0.0308, 0.0208, 0.0280, 0.350\}\\
    \hline 
    $\bar{P}^{\text{MS,dis}}$ & \{0.0197, 0.018, 0.0189, 0.0174\}\\
    \hline
    $\bar{P}^{\text{MS,ch}}$ & \{0.0185, 0.0125, 0.0168, 0.210\}\\
    \hline 
\end{tabular}
\end{table}

\begin{table} [!tb]
 \caption{Parameters of the Wildfire Simulation}
 \label{Tab:WF parameters}
 \centering
 \begin{tabular}{c|c}
    \hline
    \textbf{Parameter} & \qquad \textbf{Value}\\  
    \hline
    $\rho^{\text{b}}, H, \varphi $ & 40, 15, 200 \\
    \hline
    $\tau^{\text{a}}, \sigma^{\text{f}}, \kappa_{\text{B}}, T_{\text{f}}$ & 1, 1, $5.67\times 10^{-8}$, 1400\\
    \hline
    $\varrho, \epsilon, \zeta, \sigma^{\text{c}}$ & $2.1\times 10^{-2}$, 18,  $4.04\times 10^{-3}$, 0.78\\
    \hline
    $R_{ij,ref}, T_{ij,ref}, \bar T_{ij} $ & 15, 293, 350\\
    \hline
    $\mu^{\text{a}}, \rho^{\text{a}}, \kappa^{\text{a}}$ & $2.624\times10^{-2}, 1.184, 1.849\times10^{-5}$\\
    \hline
    
\end{tabular}
\end{table}

\subsection{Wildfire Simulation and DLR Effect}

The parameters of the active wildfire simulation are given in Table \ref{Tab:WF parameters}. Based on our simulation, any power line is out of service when the line conductor temperature $T_{ij}$ exceeds \SI{350}{K}.
We evaluate the effect of separating the DLR constraints from the optimization problem by checking the change of the binary indicator variables. 
Let $u^{\text{inc}}$ denote the optimal solution of the line outage indicator to the problem \eqref{OptFinal} with DLR. In contrast, $u^{\text{exc}}$ is the counterpart obtained by independently solving \eqref{eq:DRL}, where \eqref{Eqtn:resistive heat gain} is replaced by \eqref{cnst:DLRdetach}.
The performance metric is the percentage of the mean absolute difference (PMAD) defined as
\begin{align*}
    \text{PMAD} =  \frac{1}{N_{s} T} {\sum_{s=1}^{N_{s}}} {\sum_{t=1}^{T}} \left|u^{\text{inc}}_{s,t}- u^{\text{exc}}_{s,t}\right| \times 100\%.
    \label{eq:PMAD}
\end{align*}
As we can see in Table \ref{Tab:line outage diff}, the difference is quite small, which decreases with the increased number of scenarios. Thus, this result corroborates the validity of our proposed method. 

\begin{table} [t!]
 \caption{The PMAD of the tie line outage indicator including vs excluding DLR in the optimization problem}
 \label{Tab:line outage diff}
 \centering
 \begin{tabular}{c|c|c|c|c}
    \hline
   $N_s$ & 10 & 25 & 40 & 50\\  
    \hline  
    PMAD & 4\% & 3.2\% & 2.8\% & 2.6\%  \\
    \hline
\end{tabular}
\end{table}

Table \ref{tab:other effects} compares the effect of fire spreading factors on load shedding. Factors working against fire growth postpone the line outage time, which reduces the amount of load curtailment. For example, increased firefighting efforts ($\text{F}$) and the presence of water bodies or downhill terrain reduce the amount of curtailed load.

\begin{table}[!tb]
    \centering

        \caption{Effects of fire spreading factors in load curtailment}
        \begin{tabular}{c|c|c}
               \hline
               \textbf{Factors} & \textbf{Value} & \textbf{Curtailed load (MW)}\\
               \hline
                &0 (null effort)& 2.908\\
               Human intervention $(\text{F})$ &0.6 (moderate) & 2.075\\
               & 0.9 (extensive) & 1.615\\
               \hline
               Natural & 0 (not existing) & 2.88 \\
               obstacle $(\text{N})$ & 1.2 (existing) & 0.53\\
               \hline
               Topography $(\text{M})$ & 0 (uphill) & 2.469\\
                & 1 (downhill) & 1.83\\
               \hline
        \end{tabular}
        \label{tab:other effects}
    \end{table}

\subsection{Smoke Effect on PV Generation}
\label{Smoke effect data}
The PV output data of the Bennion Creek fire in Utah were collected from \cite{solar_data}. The PM2.5 and PM10 data for the same location are from \cite{PM_data} while the other meteorological data can be found in \cite{met_data}. 
Based on this data set, the optimal coefficients of the regression model \eqref{Eqtn:smoke_effect} are given in Table \ref{Tab:coeff for smoke effect}. Clearly, DNI is a leading factor that contributes the most to the PV output, while temperature also has a positive correlation. Relative humidity, cloud opacity, PM2.5, and PM10 exhibit negative correlations with the response, indicating an inverse relationship. This regression result confirms our intuition of how those features affect the PV generation.  

\begin{table}[!t]
 \caption{Learned coefficients of the features for the smoke effect}
 \label{Tab:coeff for smoke effect}
 \centering
 \begin{tabular}{l|l}
    \hline
    \textbf{Variable} & \textbf{Coefficient}\\  
    \hline
    Constant term & $\alpha_0 = 47.102$\\
    \hline
    Temperature ($Z_{1}$) & $\alpha_1=27.9705$\\
    \hline
    Relative humidity ($Z_{2}$) & $\alpha_2=-4.832$\\
    \hline
    DNI ($Z_{3}$) & $\alpha_4=32.377$\\
    \hline
    Cloud opacity ($Z_{4}$) & $\alpha_4=-2.037$\\
    \hline
    PM2.5 ($Z_{5}$) & $\alpha_5=-0.603$\\
    \hline
    PM10 ($Z_{6}$) & $\alpha_6=-7.857$\\
    \hline
\end{tabular}
\end{table}

\subsection{MG Operation}\label{MG_operation}

\begin{figure}[!t]
	\centering
	\includegraphics[width=0.9\linewidth]{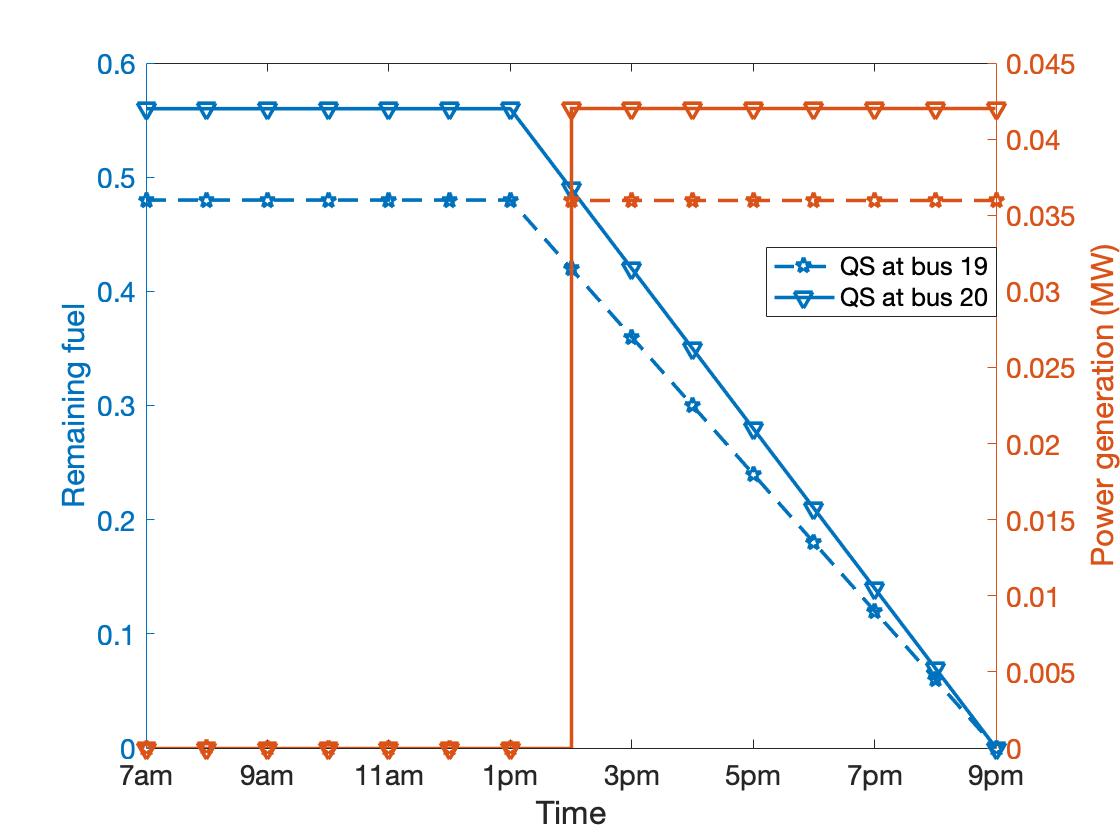}
	\caption{Fuel expenditure profiles for the QS units.}
	\label{fig:fuel_and_PG}
\end{figure}

\begin{figure}[!tb]
	\centering
	\includegraphics[width=0.9\linewidth]{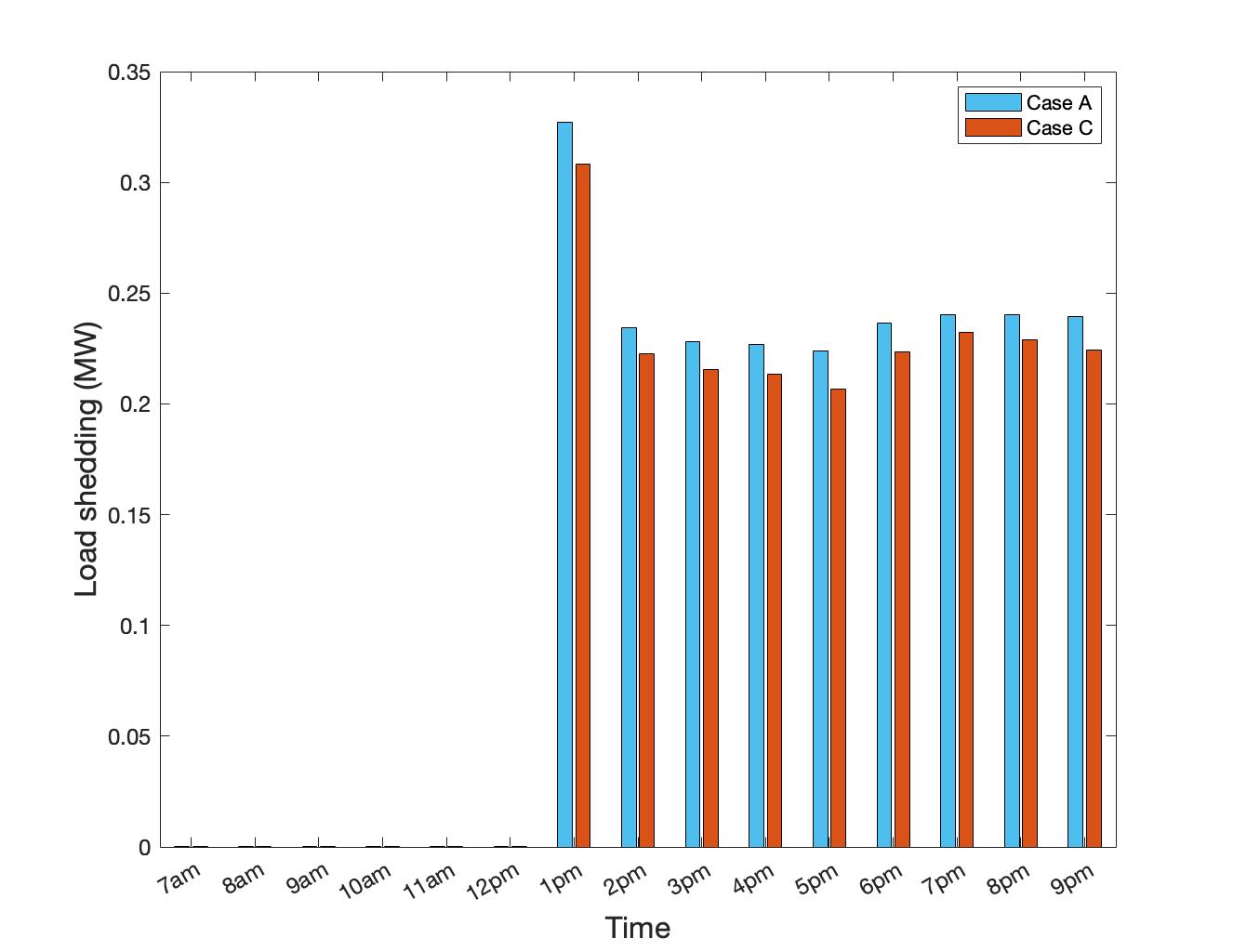}
	\caption{Total amount of load shedding considering smoke effect (Case C) and without considering smoke effect (Case A).}
	\label{fig:ls_dif.png}
\end{figure}

Figure \ref{fig:fuel_and_PG} to figure \ref{fig:ls_dif.png} are carried down for case I, with moderate firefighting effort, without any natural obstacle and a flat landscape. Figure \ref{fig:fuel_and_PG} shows the fuel consumption profile for the two QS units. As we can see in the figure, before the line outage takes place at 1 pm, the fuel levels remain the same as their initially allocated amounts (cf. Table~\ref{Tab:reserve amount}). Then, the QS generators start consuming fuels to support the load demand from 2 pm.

\begin{table}[!t]
 \caption{Fuel allocation for QS units}
 \label{Tab:reserve amount}
 \centering
\begin{tabular}{*5c} 
 \cline{1-5}
  &\multicolumn{2}{|c}{\textbf{Without smoke effect}} & \multicolumn{2}{?c}{\textbf{With smoke effect}}\\
 \hline 
Allocated fuels &\multicolumn{1}{|c}{QS unit 1} & \multicolumn{1}{c}{QS unit 2} & \multicolumn{1}{?c}{QS unit 1} & \multicolumn{1}{c}{QS unit 2}\\
  $X_i$ &\multicolumn{1}{|c}{0.4000} & \multicolumn{1}{c}{0.4400} & \multicolumn{1}{?c}{0.4799} & \multicolumn{1}{c}{0.5599}\\
\hline
\end{tabular}
\end{table}

\begin{figure}[!t]
	\centering
	\includegraphics[width=0.9\linewidth]{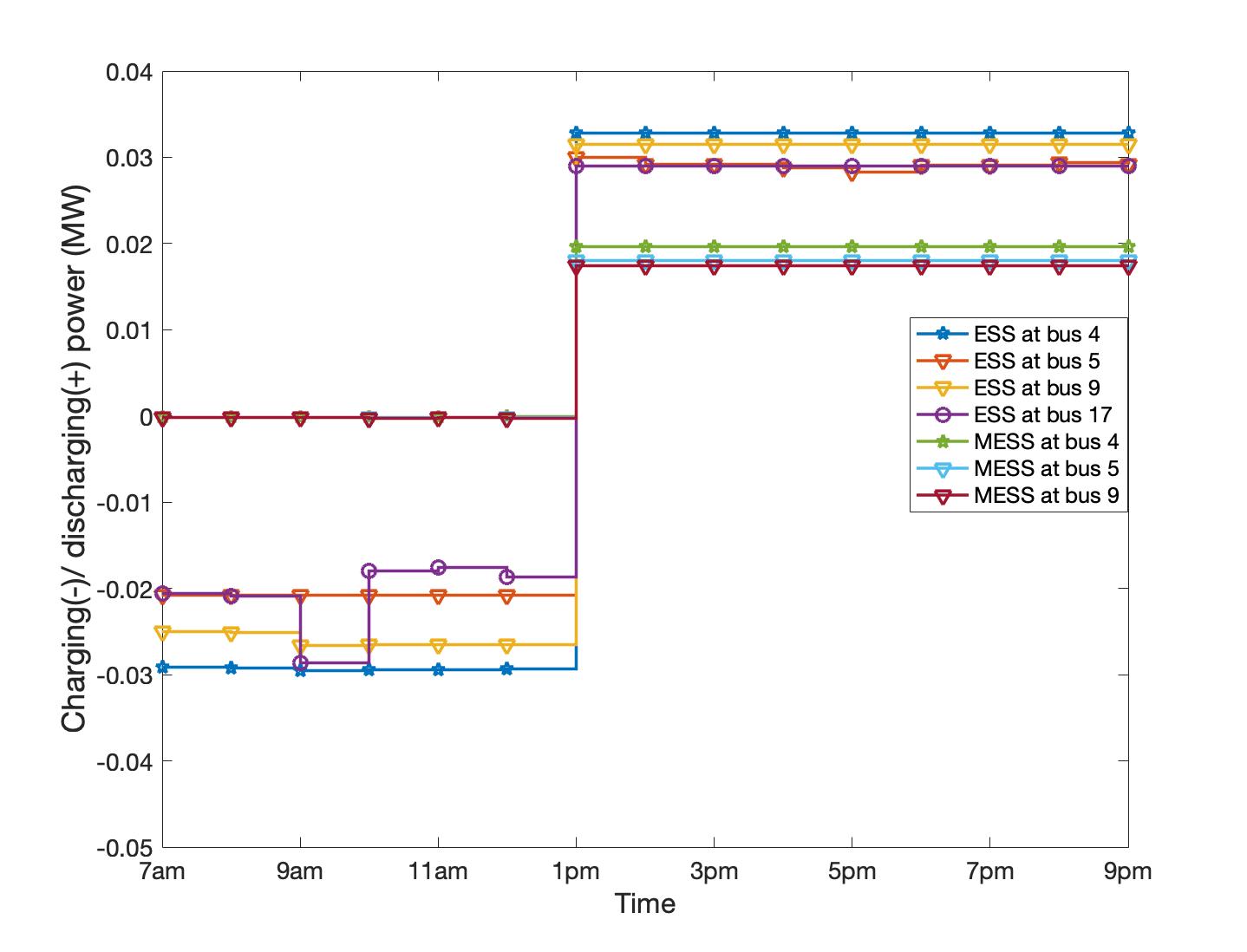}
	\caption{Optimal scheduling of ES and MS charging/discharging power.}
	\label{fig:pess}
\end{figure}

Figure \ref{fig:pess} shows the charging and discharging power scheduling of all four ES and three MS units. The ES units keep charging before islanding and discharging afterward to prioritize load serving. As the MS units were fully charged at the time of siting, they start discharging as soon as the line gets tripped off. Clearly, the limits of charging/discharging power and SoC are always respected during the entire time horizon.

Three MS units are sited at bus 4,5 and 9. Table \ref{tab:MS units} shows that without allocating these resources, the total amount of load shed is higher than having them at the designated buses.

\begin{table}[!tb]
\centering

        \caption{Impact of allocating MS units}
        \begin{tabular}{c|c|c}
        \hline
        \textbf{Sited bus index} & \textbf{Load shed with MS} &\textbf{Load shed w/o MS}\\
        \hline
        4, 5, 9 & 2.075 (MW) & 2.5694 (MW)\\
        \hline
        \end{tabular}
        \label{tab:MS units}
    \end{table}

\begin{figure}[!t]
	\centering
	\includegraphics[width=0.9\linewidth]{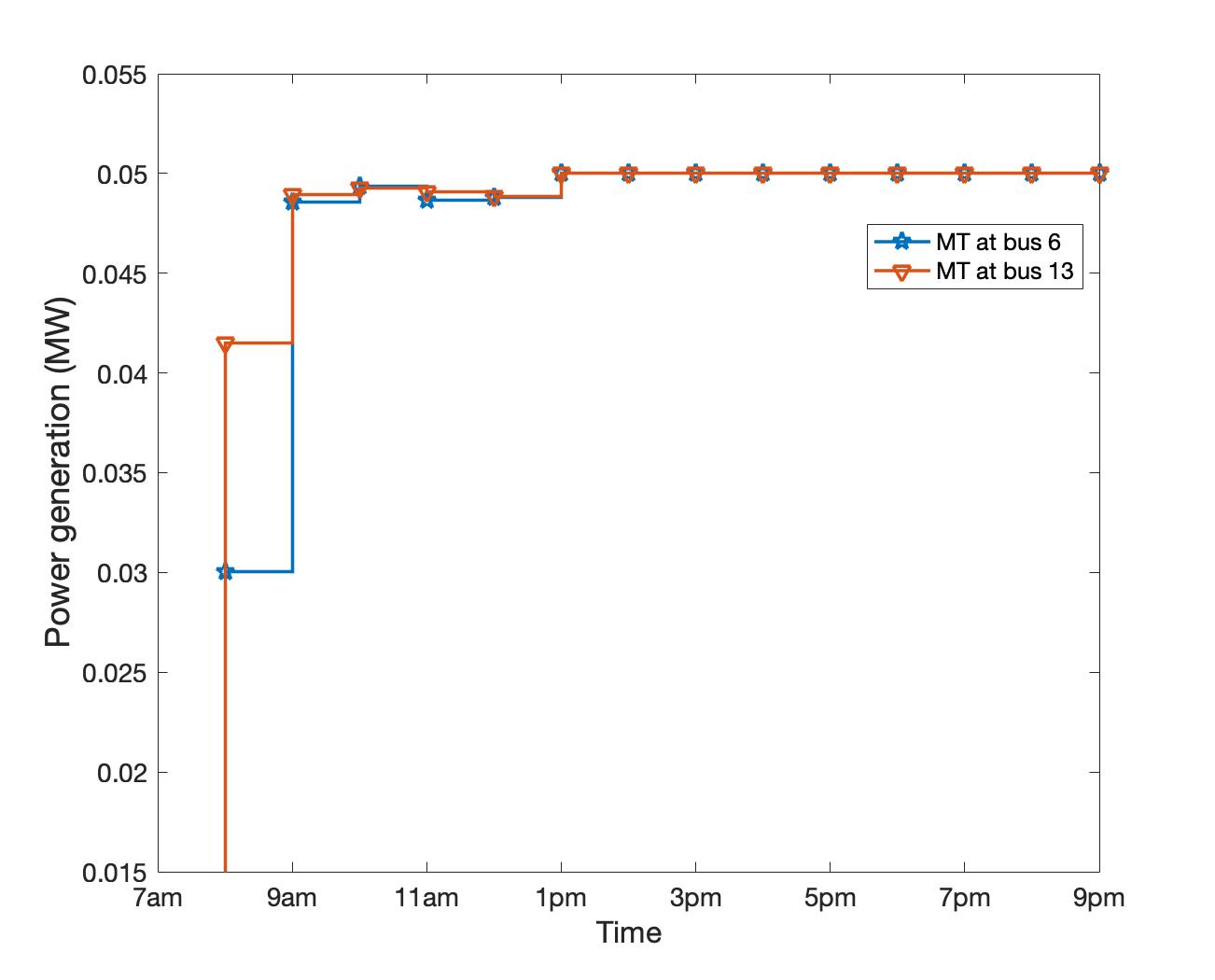}
	\caption{Optimal real power generation by the MTs.}
	\label{fig:MT_power.png}
\end{figure}

Figure \ref{fig:MT_power.png} depicts the real power generated by the two MTs. Before islanding, the MG mainly relies on purchased power from the upstream grid, which is cheaper than utilizing the MT units. Hence, the initial power generation of these units is low. After the tie line outage, the MTs ramp up their generation to support the critical loads without upstream power.

\begin{figure}[t!]
	\centering
	\includegraphics[width=0.9\linewidth]{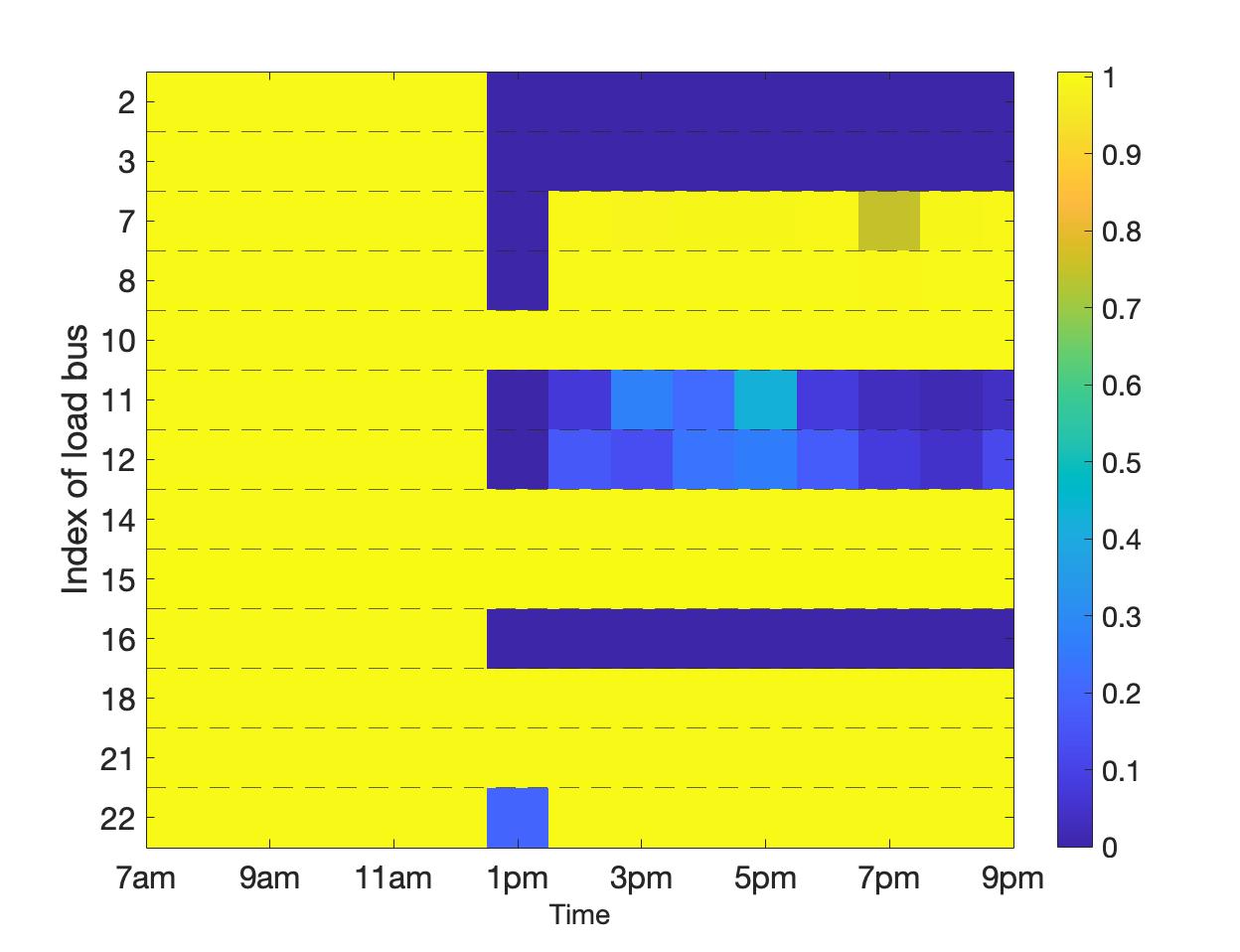}
	\caption{Serving status ($\alpha^{\text{ls}}$) of each load bus for case I.}
	\label{fig:priority_load_pu.png}
\end{figure}

Figure \ref{fig:priority_load_pu.png} shows the load serving status of each bus prioritizing the critical loads. The color gradient indicates the fraction of the picked-up demand $\alpha^{\text{ls}}$. According to the criticality factor given in Table \ref{Tab:Network parameters}, the least critical loads are located at bus 2, 3, 7, 8, 11, 12 and 16. Hence, they are shed as soon as the MG becomes islanded and all other loads are instantly supported by the distributed resources. Then, the higher priority loads among them i.e. 7 and 8 are gradually picked up by the QS units.

To show the effectiveness of our proposed approach, we compare it with two other cases. Case A is MG operation without considering the smoke effect. Case B is the one including the smoke effect but no provision of QS units. case C is our proposed one (including both smoke effect and QS units). Figure \ref{fig:ls_dif.png} displays the benefit of incorporating the smoke effect. Without considering the smoke effect, 
The PV generation is overestimated, leading to a reduced purchase of reserves (cf. Table \ref{Tab:reserve amount}). The reserves are insufficient to compensate for the reduced generation during an actual smoky scenario. In Figure \ref{fig:ls_dif.png}, it is demonstrated that the total amount of load shedding is higher in such a case. The maximum amount of load shedding occurs right at 1 pm when the islanding happens. From 2 pm, the QS units start reducing the amount of unserved loads. Clearly, there is no load shedding for the first 6 hours since all MG loads can be supported by the power from the upstream grid.

\begin{table} [!tb]
\centering
 \caption{The cost comparison between case B and C (unit in \$)}
 \label{Tab:cost comparison between B and C}
 \begin{tabular}{c|c|c}
    \hline
    \textbf{Type of cost} & \textbf{Case B} & \textbf{Case C}\\ 
    \hline
    MT generation cost & \$96.6531 & \$96.6531\\
    \hline
    QS generation cost & -- & \$248.5596\\
    \hline
    Load shedding cost & \$72,507.67 & \$48,079.80\\
    \hline
    Total cost  & \$72,604.32 & \$48,425.01\\
    \hline 
\end{tabular}
\end{table}

To compare cases B and C, we calculate the total generation cost and load shedding cost with and without the QS units. Note that both of these cases consider the smoke effect. According to Table \ref{Tab:cost comparison between B and C}, the generation cost in our proposed method (case C) is higher because it involves both MT and QS units. 
However, in case B we have to shed a lot of load due to insufficient generation. This inevitably increases the total cost.

\begin{figure}[!tb]
	\centering
	\includegraphics[width=0.9\linewidth]{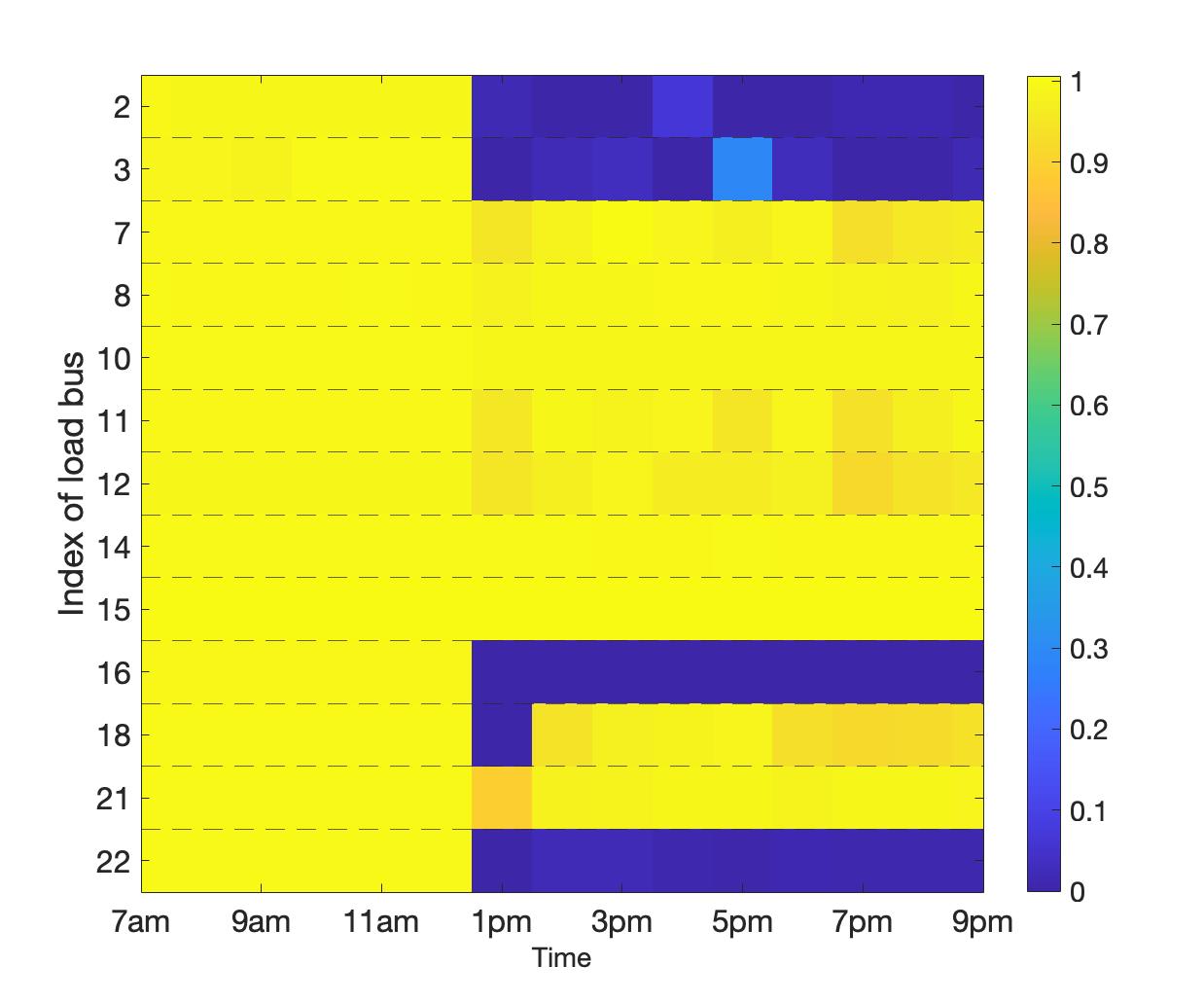}
	\caption{Serving status ($\alpha^{\text{ls}}$) of each load bus for case II.}
	\label{fig:pu_second_outage_line.png}
\end{figure}

Now, we consider another case (case II) where a leftward fire affects the line connecting bus 14 and 16, see figure \ref{fig:outage_lines}. The load serving status is shown in figure \ref{fig:pu_second_outage_line.png}. Bus 16,18,21 and 22 are disconnected from the upper sub-network (bus 1 to bus 15). In the lower network, the lowest priority load at bus 22 gets curtailed due to insufficient resources. On the other hand, resources in the upper network can be fully utilized to serve comparatively low priority loads (load at bus 11 and 12), which are unserved in case I. Table \ref{tab:line outage comparison} shows that the total load shed amount is lower with a higher generation cost in case I while case II experiences the opposite.

\begin{table}[!tb]
\centering
        \caption{Comparison between different line outages}
        \begin{tabular}{c|c|c}
               \hline
                \textbf{Criterion} & \textbf{Case I} & \textbf{Case II} \\
               \hline
                Buses between the outaged lines & 1, 2 & 14, 16\\
                \hline
               Total load shed (MW) & 2.075 & 2.163  \\
               \hline
               Generation cost (\$)& 345.21 & 325.96\\
               \hline
        \end{tabular}
        \label{tab:line outage comparison}
    \end{table}

\begin{figure}[!tb]
	\centering
	\includegraphics[width=0.95\linewidth]{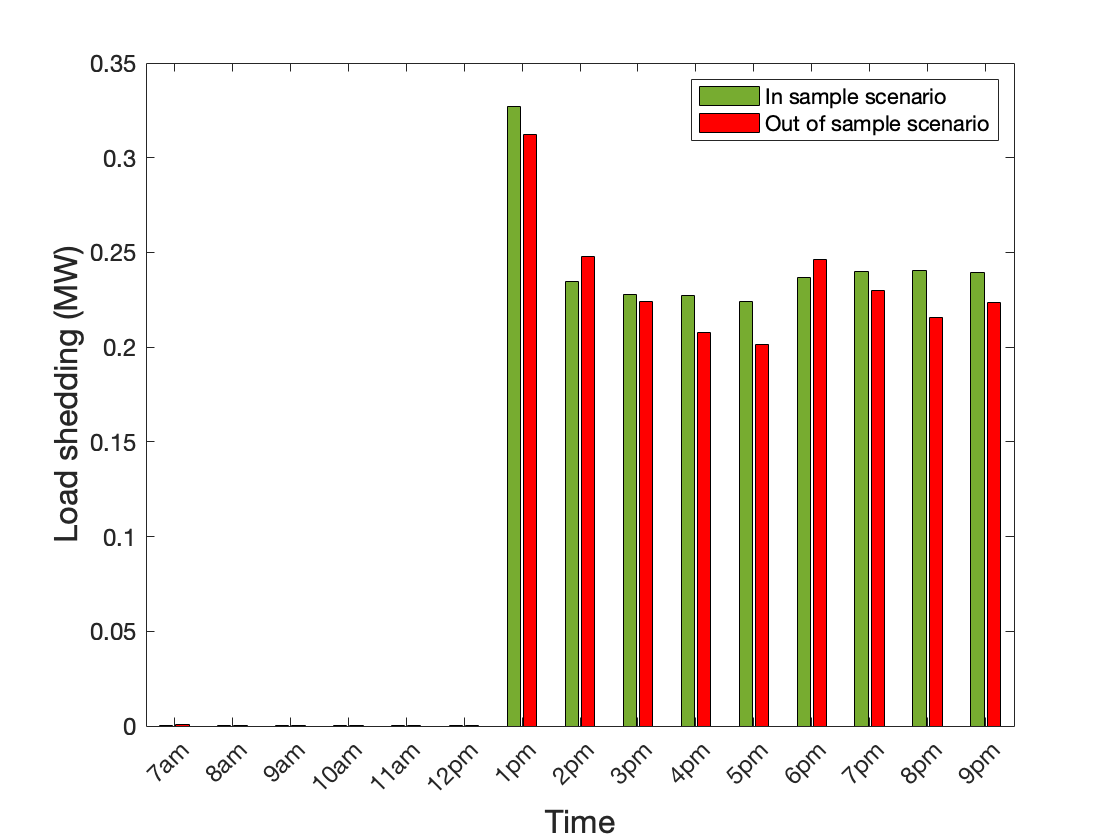}
	\caption{The comparison of load shedding in-sample and out-of-sample cases.}
	\label{fig:ls_oos.png}
\end{figure}

Figure \ref{fig:ls_oos.png} demonstrates the comparison of total load shedding between an in-sample and an out-of-sample scenario. Depending on the variation of renewable generation and active power demand, the amount of unserved load sometimes exceeds the in-sample case, and sometimes remains lower. Nevertheless, the difference in any time step is not too big, which shows the robustness of the optimal solution to our two-stage stochastic program.

Finally, Table \ref{Tab:scenario comparison} lists the optimal objective values and the computation time for solving the stochastic program. With the increased number of scenarios, the probability of simulating the actual scenario gets higher. By trading off between the computation time and optimality, we choose 50 as the number of scenarios.

\begin{table} [!tb]
 \caption{Comparison among different number of scenarios}
 \label{Tab:scenario comparison}
 \centering
 \begin{tabular}{c|c|c}
    \hline
    \textbf{Number of scenarios} & \textbf{Optimal value} & \textbf{Computation time }\\ 
    & ($\times 10^{3}$) & (sec)\\
    \hline
    10 & -478.593 & 204.30\\
    \hline
    25 & -465.672 & 678.98\\
    \hline
    40 & -466.566 & 1432.87\\
    \hline
    50 & -465.789 & 2550.567\\
    \hline
\end{tabular}
\end{table}

\section{Conclusion}

In this work, we formulate a two-stage stochastic OPF for operating an MG during a wildfire. A comprehensive model of fire dynamics causing line outages and altered PV generation stemming from the smoke plume is analyzed. To prepare for the upcoming contingency more accurately while minimizing load shedding, 
we utilize renewable fuel-powered quick-start generators and mobile energy storage devices. 
The proposed framework significantly speeds up computations by effectively separating 
the DLR constraints from the OPF problem. The extensive simulations demonstrate the merit of the proposed model by comparing the power outage and operational costs in various case studies. Without considering the smoke effect, PV generation is overestimated, which results in increased load shedding and negative impacts on the system's resilience. In the future, we will utilize advanced sensors and communication technologies to characterize the uncertainties of fire progression more accurately. Moreover, incorporating wind propagation factors can improve the modeling of smoke effects on actual PV generation. Another compelling direction is to harness clean energy technologies to enhance grid sustainability and resilience.

\bibliographystyle{IEEEtran}

\begin{IEEEbiography}[{\includegraphics[width=1in,height=1.25in,clip,keepaspectratio]{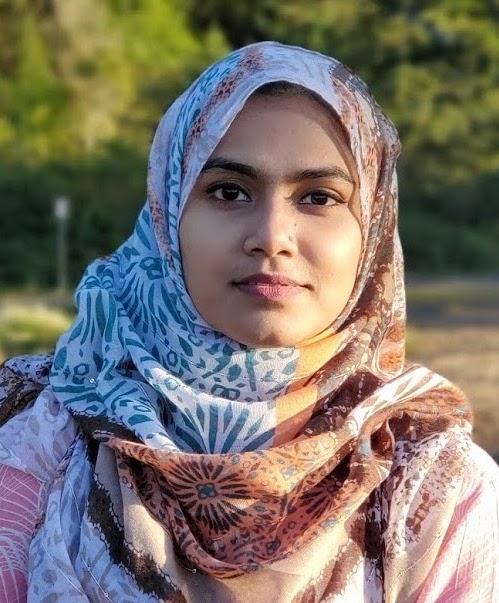}}]{Sifat Chowdhury} received the B.S. degree in Electrical and Electronic Engineering from Bangladesh University of Engineering and Technology (BUET). She got her M.S. degree in Electrical and Computer Engineering from the University of California, Santa Cruz.
Her research interests include electric power system optimization, microgrid operation, data analytics, and enhancing grid resilience during extreme weather events.

\end{IEEEbiography}

\begin{IEEEbiography}[{\includegraphics[width=1in,height=1.25in,clip,keepaspectratio]{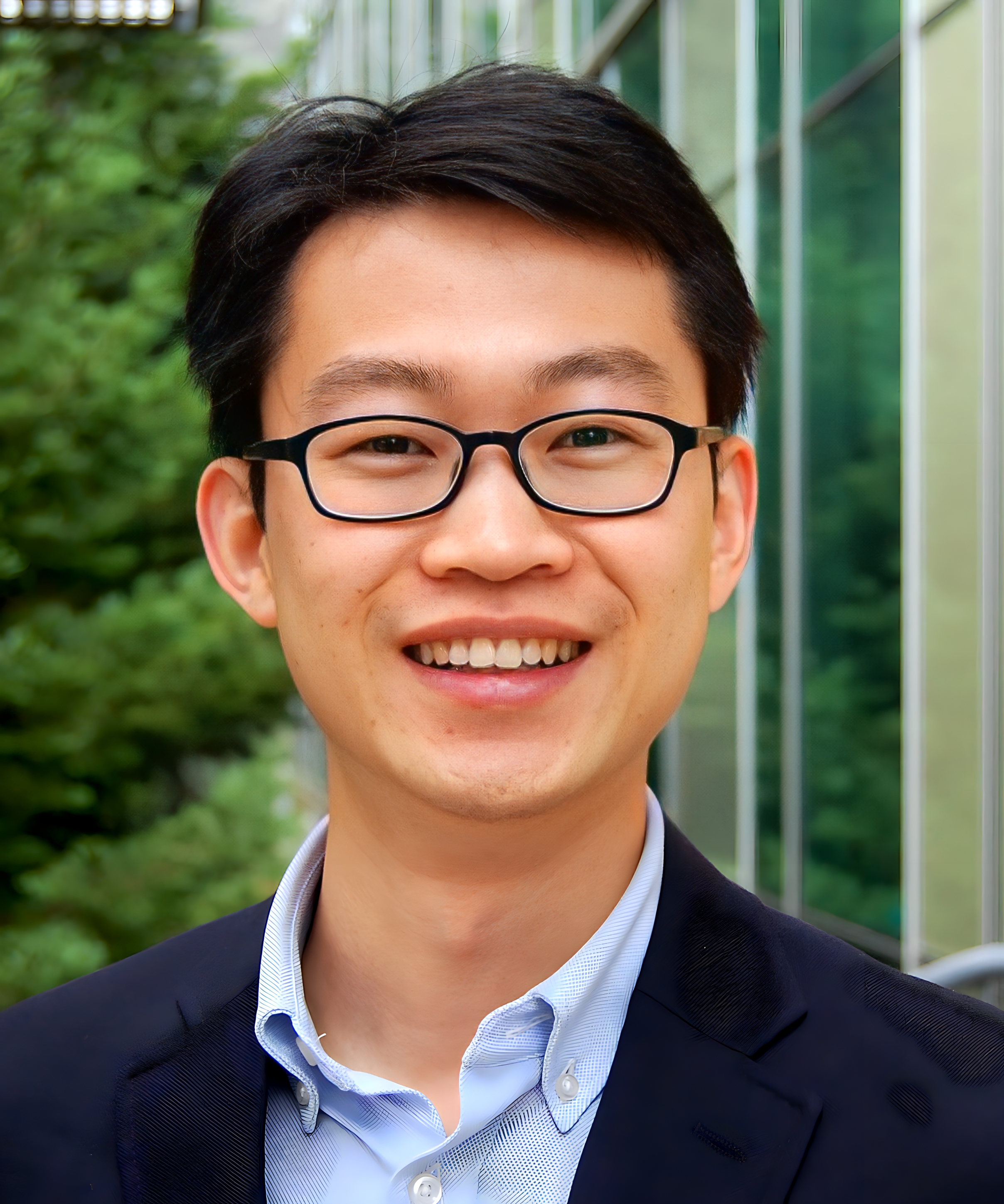}}]{Yu Zhang} (Member, IEEE) received the Ph.D.
degree in Electrical and Computer Engineering
from the University of Minnesota, Minneapolis,
MN, USA, in 2015. He is currently an Assistant Professor with the ECE Department in the University of California at Santa Cruz (UCSC). Before joining UCSC, he held a postdoctoral position with the University of California at Berkeley
and the Lawrence Berkeley National Laboratory. His research interests include cyber-physical systems, smart power grids, optimization theory, machine learning, and big data analytics. He was
a recipient of the Hellman Fellowship, in 2019. He was a co-recipient of
the Early Career Best Paper Award from the Energy, Natural Resources,
and the Environment (ENRE) Section of the Institute of Operations Research
and the Management Sciences (INFORMS), in 2021.

\end{IEEEbiography}

\EOD

\end{document}